\documentclass[11pt]{article}

\usepackage{amsmath,amsfonts,amssymb,graphics,graphicx,fancyhdr,subfigure,float,fancybox,url}
\usepackage{wrapfig}
\usepackage{cite}
\usepackage{fullpage}
\usepackage{rotating}
\usepackage{multirow} 
\usepackage{booktabs} 
\usepackage{graphicx} 
\usepackage[scaled]{helvet}
\usepackage[T1]{fontenc}
\usepackage{color} 
\usepackage[linktoc=all]{hyperref}

\newcommand{\p}{{\rm Pr}}
\newcommand{\e}{{\rm E}}

\newcommand{\var}{{\rm Var}}

\newcommand{\no}{\noindent}
\newcommand{\bh}{\widehat}
\newcommand{\bt}{\widetilde}
\newcommand{\btil}{\bt}
\newcommand{\cn}{\cite}
\newcommand{\real}{\mathbb{R}}

\newcommand{\iid}{\stackrel{i.i.d.}{\sim}}
\newcommand{ \bm }[1]{ \mbox{\boldmath ${#1}$} }

\newcommand{ \bmm }[1]{ \mbox{\boldmath ${#1}$} }
\newcommand{\bs}{\mathbf{s}}

\newcommand{\subsub}[1]{\medskip \noindent {\em {#1}}}
\newcommand{\fst}{{\rm F}_{\rm ST}}

\newcommand{\bF}{\bm{{\rm F}}}
\newcommand{\bL}{\bm{{\rm L}}}
\newcommand{\bA}{\bm{{\rm A}}}
\newcommand{\bH}{\bm{{\rm H}}}
\newcommand{\bP}{\bm{{\rm P}}}
\newcommand{\bQ}{\bm{{\rm Q}}}
\newcommand{\bS}{\bm{{\rm S}}}
\newcommand{\bV}{\bm{{\rm V}}}
\newcommand{\bU}{\bm{{\rm U}}}
\newcommand{\bD}{\bmm{\Delta}}
\newcommand{\bW}{\bm{{\rm W}}}
\newcommand{\bT}{\bm{{\rm T}}}
\newcommand{\bLambda}{\bm{\Lambda}}
\newcommand{\bx}{\bm{{\rm x}}}
\newcommand{\bZ}{\bmm{Z}}
\newcommand{\bz}{\bmm{z}}
\newcommand{\bG}{\bmm{\Gamma}}
\newcommand{\logit}{{\rm logit}}
\newcommand{\bgamma}{\bmm{\gamma}}
\newcommand{\bX}{\bm{{\rm X}}}
\newcommand{\piij}{\pi_{ij}}

\begin{document}

\title{Probabilistic models of genetic variation in structured populations applied to global human studies}
 \author{
Wei Hao$^{1 \ast}$, Minsun Song$^{1 \ast + }$, and John D. Storey$^{1,2 \; \dagger}$
\\ \footnotesize 1. Lewis-Sigler Institute for Integrative Genomics, Princeton University, Princeton, NJ 08544
\\ \footnotesize 2. Department of Molecular Biology, Princeton University, Princeton, NJ 08544
\\ \footnotesize $^{\ast}$ These authors contributed equally to this work
\\ \footnotesize $^{+}$ Present address:  Division of Cancer Epidemiology and Genetics, National Cancer Institute,  \\ \footnotesize National Institutes of Health, Rockville, MD 20850 \ \ \ \ 
\\ \footnotesize $^\dagger$ To whom correspondence should be addressed: {\tt jstorey@princeton.edu}
}
\date{ }

\maketitle

\tableofcontents

\clearpage
\section*{{\sc Abstract}} \addcontentsline{toc}{section}{\sc Abstract}
Modern population genetics studies typically involve genome-wide genotyping of individuals from a diverse network of ancestries.  An important problem is how to formulate and estimate probabilistic models of observed genotypes that account for complex population structure.  The most prominent work on this problem has focused on estimating a model of admixture proportions of ancestral populations for each individual.  Here, we instead focus on modeling variation of the genotypes without requiring a higher-level admixture interpretation. We formulate two general probabilistic models, and we propose computationally efficient algorithms to estimate them.  First, we show how principal component analysis (PCA) can be utilized to estimate a general model that includes the well-known Pritchard-Stephens-Donnelly admixture model as a special case.  Noting some drawbacks of this approach, we introduce a new ``logistic factor analysis'' (LFA) framework that seeks to directly model the logit transformation of probabilities underlying observed genotypes in terms of latent variables that capture population structure. We demonstrate these advances on data from the Human Genome Diversity Panel and 1000 Genomes Project, where we are able to identify SNPs that are highly differentiated with respect to structure while making minimal modeling assumptions.

\section{{\sc Introduction}} 
Understanding genome-wide genetic variation among individuals is one of the primary goals of modern human genetics.  Genome-wide association studies aim to identify genetic variants throughout the entire genome that are associated with a complex trait \cn{mccarthy2008,Frazer2009,Consortium2007}.  One of the major challenges in analyzing these studies is the problem of spurious associations due to population structure \cn{Pritchard1999}, and methods to deal with this are still in development \cn{astle2009,Price2010,Kang2010}.  A related effort is underway to provide a comprehensive, genome-wide understanding of how genetic variation among humans is driven by evolutionary and demographic forces \cn{jorde2001}. A rigorous characterization of this variation will lead to a better understanding of the history of migration, expand our ability to identify signatures of natural selection, and provide important insights into the mechanisms of human disease \cn{nielsen2007,Rosenberg2002}.  For example, the Human Genome Diversity Project (HGDP) is an international project that has genotyped a large collection of DNA samples from individuals distributed around the world, aiming to assess worldwide genetic diversity at the genomic level \cn{Cann2002,Rosenberg2002,Rosenberg2005}. The 1000 Genomes Project (TGP) is comprehensively cataloging human genetic variation by producing complete genome sequences of well over 1000 individuals of diverse ancestries \cn{Consortium2010}.  

Systematically characterizing genome-wide patterns of genetic variation is difficult due to the numerous and complex forces driving variation. There is a fundamental need to provide probabilistic models of observed genotypes in the presence of complex population structure.   A series of influential publications have proposed methods to estimate a model of admixture, where the primary focus is on the admixture proportions themselves \cn{PritchardStephens2000,Tang2005,Alexander:2009p2792}, which in turn may produce estimates of the allele frequencies of every genetic marker for each individual.  Here, we instead focus directly on these individual-specific allele frequencies, which gives us potential advantages in terms of accuracy and computational efficiency.  

We propose two flexible genome-wide models of individual-specific allele frequencies as well as methods to estimate them.  First, we develop a model that includes as special cases the aforementioned models; specifically, the Balding-Nichols (BN) model \cn{Balding1995} and its extension to the Pritchard-Stephens-Donnelly (PSD) model \cn{PritchardStephens2000}.  However, we identify some limitations of our method to estimate this model.  We therefore propose an alternative model based on the log-likelihood of the data that allows for rapid estimation of allele frequencies while maintaining a valid probabilistic model of genotypes.

The estimate of the first model is based on principal component analysis (PCA), which is a tool often applied to genome-wide data of genetic variation in order to uncover structure. One of the earliest applications of PCA to population genetic data was carried out by Menozzi et al.  \cn{Menozzi1978}. Exploratory analysis of complex population structure with PCA has been thoroughly studied \cn{Menozzi1978,Sokal1999a,Rendine1999,Novembre2008,Manni2010}.   We show that a particular application of PCA can also be used to estimate allele frequencies in highly structured populations, although we have to deal with the fact that PCA is a real-valued operation and is not guaranteed to produce allele frequency estimates that lie in the unit interval [0,1].   

The estimate of the second model is based on a generalized factor analysis approaches that directly model latent structure in observed data, including categorical data \cn{BKM2011} in which genotypes are included.  We utilize a factor model of population structure \cn{Engelhardt2010} in terms of nonparametric latent variables, and we propose a method called ``logistic factor analysis'' (LFA) that extends the PCA perspective towards likelihood-based probabilistic models and statistical inference.  LFA is shown to provide accurate and interpretable estimates of individual-specific allele frequencies for a wide range of population structures.  At the same time, this proposed approach provides visualizations and numerical summaries of structure similar to that of PCA, building a convenient bridge from exploratory data analysis to probabilistic modeling.  

We compare our proposed methods to existing algorithms (ADMIXTURE \cn{Alexander:2009p2792} and fastStructure \cn{Raj2013}) and show that when the goal is to estimate all individual-specific allele frequencies, our proposed approaches are conclusively superior in both accuracy and computational speed.  We apply the proposed methods to the HGDP and TGP data sets, which allows us to estimate allele frequencies of every SNP in an individual-specific manner.  Using LFA, we are also able to rank SNPs for differentiation according to population structure based on the likelihoods of the fitted models.  In both data sets, the most differentiated SNP is proximal to {\em SLC24A5}, and the second most differentiated SNP is proximal to {\em EDAR}.  Variation in both of these genes has been hypothesized to be under positive selection in humans.  In the TGP data set, the second most different SNP is rs3827760, which confers a missense mutation in {\em EDAR} and has been recently experimentally validated as having a functional role in determining a phenotype \cn{Kamberov:2013p2731}.  We also identify several SNPs that are highly differentiated in these global human studies that have recently been associated with diseases such as cancer, obesity, and asthma.  

\section{{\sc Methods}} 
\subsection{Models of Allele Frequencies}
\label{allelefreqs}
\begin{figure}[!h]
\centerline{\includegraphics[width=0.7\textwidth]{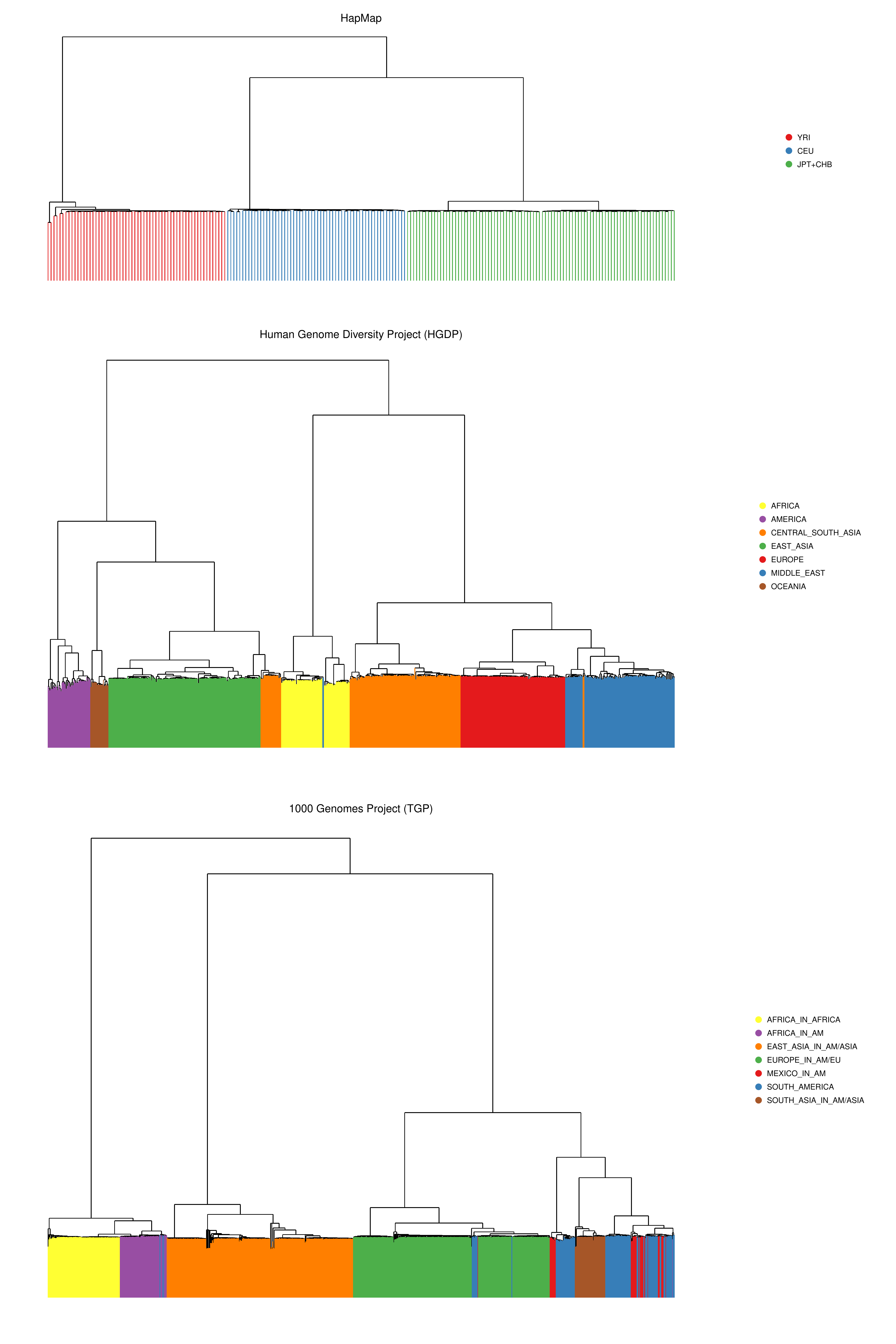}}
\caption{A hierarchical clustering of individuals from the HapMap, HGDP, and TGP data sets.  A dendrogram was drawn from a hierarchical clustering using Ward distance based on SNP genotypes (MAF $> 5\%$). Whereas the HapMap project shows a definitive discrete population structure (by sampling design), the HGDP  and TGP data show the complex structure of human populations.}
\label{fig:cluster}
\end{figure}

It is often the case that human and other outbred populations are ``structured'' in the sense that the genotype frequencies at a particular locus are not homogeneous throughout the population \cn{astle2009}. Geographic characterizations of ancestry often explain differing genotype frequencies among subpopulations. For example, an individual of European ancestry may receive a particular genotype according to a probability different than an individual of Asian ancestry. This phenomenon has been observed not only across continents, but on very fine scales of geographic characterizations of ancestry.  Recent studies have shown that population structure in human populations is quite complex, occurring more on a continuous rather than a discrete basis \cn{Rosenberg2002}. We can illustrate the spectrum of structural complexity with Figure \ref{fig:cluster}, which shows dendrograms of hierarchically clustered individuals from the HapMap (phase II), HGDP, and TGP data sets.  The HapMap samples strongly indicate explicit membership of each individual to one of three discrete subpopulations (due to the intended sampling scheme). On the other hand, the clusterings of the HGDP and TGP individuals show a very complex configuration, more representative of random sampling of global human populations.  

Let us introduce $\bZ$ as an unobserved variable capturing an individual's structure. Let $x_{ij}$ be the observed genotype for SNP $i$ and individual $j$ ($i=1,\ldots,m$, $j=1,\ldots,n$), and assume that $x_{ij}$ is coded to take the values $0, 1, 2$.  We will call the observed $m \times n$ genotype matrix $\bX$.    For SNP $i$, the allele frequency can viewed as a function of $\bZ$, i.e. $\pi_i (\bZ)$. For a sampled individual $j$ from an overall population,  we have ``individual-specific allele frequencies'' \cn{Thornton:2012p2787} defined as $\pi_{ij} \equiv \pi_{i}(\bz_j)$ at SNP $i$. Each value of $\pi_{ij}$ informs us as to the expectation of that particular SNP/individual pair, supposing we observed a new individual at that locus with the same structure; i.e. $\e[x_{ij}]/2 = \pi_{ij}$.  If an observed SNP genotype $x_{ij}$ is treated as a random variable, then under Hardy-Weinberg Equilibrium $\pi_{ij}$ serves to model $x_{ij}$ as a Binomial parameter: $x_{ij} \sim \mbox{Binomial}(2,\pi_{ij})$.  The focus of this paper is on the simultaneously estimation of all $m \times n$ $\pi_{ij}$ values.

The flexible, accurate, and computationally efficient estimation of individual-specific allele frequencies is important for population genetic analyses.  For example, Corona et al. (2013) \cn{Corona:2013p2708} recently showed that considering the worldwide distribution of allele frequencies of SNPs known to be associated with human diseases may be a fundamental component to understanding the relationship between ancestry and disease.  Testing for Hardy-Weinberg equilibrium reduces to testing whether the genotype frequencies for SNP $i$ follow probabilities $\pi_{ij}^2$, $2\pi_{ij}(1-\pi_{ij})$, and $(1-\pi_{ij})^2$ for all individuals $j=1, \ldots, n$.  It can be shown that the well-known $\fst$ measure can be characterized for SNP $i$ using values of $\pi_{ij}$, $j=1, 2, \ldots, n$ (Section \ref{fst}). Finally, we have recently developed a test of association that corrects for population structure and involves the estimation of $\log\left(\frac{\pi_{ij}}{1-\pi_{ij}}\right)$ \cn{Song2013}.  Therefore, flexible and well-behaved estimates of the individual-specific allele frequencies $\pi_{ij}$ are needed for downstream population genetic analyses.

It is straightforward to write other models of population structure in terms of $\bZ$. For the Balding-Nichols model, each individual is assigned to a population, thus $\bz_j$ indicates individual $j$'s population assignment. For the Pritchard-Stephens-Donnelly (PSD) model, each individual is considered to be an admixture of a finite set of ancestral populations. Following the notation of \cn{PritchardStephens2000}, we can write $\bz_j$ as a vector with elements $q_{kj}$, where $k$ indexes the ancestral populations, and we constrain $q_{kj}$ to be between 0 and 1 subject to $\sum_k q_{kj} =1$. Assuming the PSD model allows us to write each $\pi_{ij} = \sum_k p_{ik} q_{kj}$ and leads to a matrix form: $\bF = \bP \bQ$, where $\bF$ is the $m \times n$ matrix of allele frequencies with $(i,j)$ entry $\pi_{ij}$, $\bP$ is the $m \times d$ matrix of ancestral population allele frequencies $p_{ik}$, and $\bQ$ is the $d \times n$ matrix of admixture proportions. The elements of $\bP$ and $\bQ$ are explicitly restricted to the range $[0,1]$. 

The PSD model is focused on the matrices $\bP$ and $\bQ$, which have standalone interpretations, but we aim instead to estimate all $\pi_{ij}$ with a high level of accuracy and computational efficiency. Writing the structure of the allele frequency matrix $\bF$ as a linear basis, we have:
\begin{equation}\label{eq:pi}
\mbox{\textbf{Model 1:}\ \ \ \ \ } \bF=\bG \bS
\end{equation}
where $\bG$ is $m \times d$ and $\bS$ is $d\times n$ with $d \leq n$. The $d\times n$ matrix $\bS$ encapsulates the genetic population structure for these individuals since $\bf S$ is not SNP-specific. The $m \times d$ matrix $\bG$ maps how the structure $\bS$ is manifested in the allele frequencies. Operationally, each SNP's allele frequency are a linear combination of the rows of $\bS$, where the linear weights for SNP $i$ are contained in row $i$ of $\bG$. We define the dimension $d$ so that $d=1$ corresponds to the case of no structure: when $d=1$, $\bS = (1, 1, \ldots, 1)$ and $\bG$ is the column vector of marginal allele frequencies. 

This model is not necessarily the most effective way to estimate $\pi_{ij}$ when working in the context of a probabilistic model or with the likelihood function given the data. Model 1 resembles linear regression, where the allele frequencies are treated as a real-valued response variable that is linearly dependent on the structure. A version of regression for the case of categorical response variables (e.g., genotypes) with underlying probability parameters is logistic regression. We developed an approach we call ``logistic factor analysis'', which is essentially an extension of nonparametric factor analysis to $\{0,1,2\}$ valued genotype data. Much of the justification for LFA is similar to that of {\em generalized linear models} \cn{GLM}.

The log-likelihood is the preferred mathematical framework for representing the information the data contain about unknown parameters \cn{TPE}. Suppose that Hardy-Weinberg equilibrium holds such that $x_{ij} \sim \mbox{Binomial}(2, \pi_{ij})$. We can write the log-likelihood of the data for SNP $i$ and individual $j$ as:
\begin{align*}
\ell(\pi_{ij} | x_{ij}) &= \log \left( \p(x_{ij} | \pi_{ij}) \right) \propto \log \left( \pi_{ij}^{x_{ij}} (1-\pi_{ij})^{2-x_{ij}} \right)\\
&= x_{ij} \log\left( \frac{\pi_{ij}}{1-\pi_{ij}} \right) + 2\log(1-\pi_{ij}).
\end{align*}
The log-likelihood of SNP $i$ for all unrelated individuals is the sum: $\sum_{j=1}^{n} \ell(\pi_{ij} | x_{ij})$. The term $\log\left( \frac{\pi_{ij}}{1-\pi_{ij}} \right)$ is the logit function and is written as $\logit(\pi_{ij})$. $\logit(\pi_{ij})$ is called the ``natural parameter''  or ``canonical parameter'' of the Binomial distribution and is the key component of logistic regression.  An immediate benefit of working with $\logit(\pi_{ij})$ is that it is real valued, which allows us to directly model $\logit(\pi_{ij})$ with a linear basis.

Let $\bL$ be the $m \times n$ matrix with $(i,j)$ entry equal to $\logit(\pi_{ij})$.  We formed the following parameterization of $\bL$:
\begin{equation} \label{eq:logitmodel}
\mbox{\textbf{Model 2:}\ \ \ \ \ } \bL=\bA \bH
\end{equation}
where $\bf A$ is $m\times d$ and $\bH$ is $d\times n$ with $d \leq n$.  In this case we can write
$$
\logit(\pi_{ij}) = \sum_{k=1}^{d} a_{ik} h_{kj}, 
$$
where all parameters are free to span the real numbers $\real$. 

We call the rows of $\bH$ ``logistic latent factors'' or just ``logistic factors'' as they represent unobserved variables that explain the inter-individual differences in allele frequencies. In other words, the $\logit$ of the vector of individual-specific allele frequencies for SNP $i$ can be written as a linear combination of the rows of $\bH$:
$$
[\logit(\pi_{i1}), \ldots, \logit(\pi_{in})] = \logit(\bm{\pi}_i) = \sum_{k=1}^{d} a_{ik} \bm{h}_k,
$$
where $\bm{h}_k$ is the $k$th row of $\bH$. Likewise, we can write:
$$
(\pi_{i1}, \ldots, \pi_{in}) = \bm{\pi}_i = \frac{\exp\left[\sum_{k=1}^{d} a_{ik} \bm{h}_k\right]}{1 + \exp\left[\sum_{k=1}^{d} a_{ik} \bm{h}_k\right]}.
$$

The relationship between our proposed LFA approach and existing approaches of estimating latent variables in categorical data is detailed in Section~\ref{existingmethods}. Specifically, it should be noted that even though we propose calling the approach ``logistic factor analysis'', we do not make any assumptions about the distribution of the factors (which are often assumed to be Normal). A technically more detailed name of the method is a ``logistic nonparametric linear latent variable model for Binomial data.''

\subsection{Estimation and Algorithms}
\label{algs}

The two models presented earlier make minimal assumptions as to the nature of the structure. For example, in Model 1, both $\bG$ nor $\bS$ are real valued. This allows us to apply an efficient PCA-based algorithm directly to the genotype matrix $\bX$, obtaining estimates of $\btil{\bF}$, $\btil{\bG}$, and $\btil{\bS}$. In essence, $\btil{\bF}$ is estimated by forming the projection of $\bX/2$ onto the top $d$ principal components of $\bX$ with an explicit intercept for the $d=1$ case. One drawback of this approach is that because PCA is designed for continuous data, we have to artificially constrain $\btil{\bF}$ to be in the range $[0,1]$. However, we show below that $\btil{\bF}$ is still an extremely accurate estimate of the allele frequencies $\bF$ for all formulations of $\bF$ considered here, including the PSD model.

\vspace{\baselineskip}
\noindent
\textbf{Algorithm 1}: Estimating $\bF$ from PCA
\begin{enumerate}
\item Let $\btil{\mu}_i$ be the sample mean of row $i$ of $\bX$.  Set $x^*_{ij} = x_{ij} - \btil{\mu}_i$ and let $\bX^*$ be the $m \times n$ matrix with $(i,j)$ entry $x^*_{ij}$.
\item Perform singular value decomposition (SVD) on $\bX^*$ which decomposes $\bX^* = \bU \bD \bV^T$.  Note that the rows of $\bD \bV^T$ are the $n$ row-wise principal components of $\bX^*$ and $\bU$ are the principal component loadings.
\item Let $\btil{\bX}^*_{d-1}$ be the projection of $\bX^*$ on the top $d-1$ eigen-vectors of this SVD, $\btil{\bX}_{d-1}^* = \bU_{1:(d-1)} \bD_{1:(d-1)} \bV_{1:(d-1)}^T$. 
\item \label{fstar} Construct $\btil{\bF}^*$ by adding $\btil{\mu}_i$ to row $i$ of $\btil{\bX}^*_{d-1}$ (for $i=1, \ldots, n$)  and multiplying the resulting matrix by $1/2$.  In mathematical terms, $\btil{\bF}^* = \btil{\bG} \btil{\bS}$ where

\begin{align*}
\btil{\bG} &= \begin{pmatrix}
\ & \frac{1}{2}\btil{\mu}_1 \\
\frac{1}{2} \bU_{1:(d-1)} \bD_{1:(d-1)} & \vdots \\
\ & \frac{1}{2}\btil{\mu}_m
\end{pmatrix} \\ 
&= \begin{pmatrix}
\frac{1}{2} u_{11} \delta_{1} & \cdots & \frac{1}{2} u_{1,d-1} \delta_{d-1} & \frac{1}{2}\btil{\mu}_1 \\
\frac{1}{2} u_{21} \delta_{1} & \cdots & \frac{1}{2} u_{2,d-1} \delta_{d-1} & \frac{1}{2}\btil{\mu}_2 \\
\vdots & \  & \vdots & \vdots  \\
\frac{1}{2} u_{m1} \delta_{1} & \cdots & \frac{1}{2} u_{m,d-1} \delta_{d-1} &  \frac{1}{2}\btil{\mu}_m
\end{pmatrix}, \\
\btil{\bS} &= \begin{pmatrix}
\bV_{1:(d-1)}^T \\
1 \; 1\; \ldots \; 1 
\end{pmatrix} \\
&= \begin{pmatrix}
v_{11} & v_{21} & \cdots & v_{n1} \\
v_{12} & v_{22} & \cdots & v_{n2} \\
\vdots & \vdots & \  & \vdots \\
v_{1,d-1} & v_{2,d-1} & \cdots & v_{n,d-1} \\
1 & 1 & \cdots & 1
\end{pmatrix},
\end{align*}
and $\delta_i$ is the $i$th diagonal entry of $\bD$. Let $\btil{\pi}^*_{ij}$ to be the $(i,j)$ entry of $\btil{\bF}^*$.  
\item Since it may be the case that some $\btil{\pi}^*_{ij}$ are such that $\btil{\pi}^*_{ ij}< 0$ or $\btil{\pi}^*_{ij} > 1$, we truncate these. The final PCA based estimate of $\bF$ is formed as $\widetilde{\bF}$ where the $(i,j)$ entry $\widetilde{\pi}_{ij}$ is defined to be
$$
\widetilde{\pi}_{ij} = \begin{cases} C & \mbox{if } \btil{\pi}^*_{ij} \leq  C
\\ \btil{\pi}^*_{ij} & \mbox{if } C < \btil{\pi}^*_{ij} < 1-C
\\ 1-C & \mbox{if } \btil{\pi}^*_{ij} \geq 1-C
\end{cases}
$$
for some $C \gtrsim 0$.  An estimate of $\bL$ can be formed as $\btil{\bL} = \logit(\btil{\bF})$.
\end{enumerate}

\no Here we used $C = \frac{1}{2n}$.  In summary, $\widetilde{\bF}$ is a projection of $\bX$ into its top principal components, scaled by $1/2$, and truncated so that all values lie in the interval $(0,1)$.

For Model 2, we propose a method for estimating the latent variables $\bH$. Starting from the $\btil{\bF}$ found by Algorithm 1, we apply the $\logit$ transformation to the subset of rows where we did not have to adjust the values that were $<0$ or $>1$, and then extract the right singular vectors of this transformed subset. As long as the subset is large enough to span the same space as the row space of $\bL$, this approach accurately estimates the basis of $\bH$. Next, we calculate the maximum likelihood estimation of $\bA$ parametrized by $\bh{\bH}$ to yield $\bh{\bA}$ and then $\bh{\bL} = \bh{\bA} \bh{\bH}$. This involves performing a logistic regression of each SNP's data on $\bh{\bH}$. In order to estimate the individual-specific allele frequency matrix $\bF$, we calculate $\bh{\bF} = \logit^{-1}(\bh{\bL})$. An important property to note is that all $\bh{\pi}_{ij} \in [0,1]$ due to the fact that we are modeling the natural parameter.

\vspace{\baselineskip}
\noindent
\textbf{Algorithm 2}: Estimating Logistic Factors
\begin{enumerate}
\item Apply Algorithm 1 to obtain the estimate $\btil{\bF}^*$ from Step \ref{fstar}.  
\item Recalling that $\btil{\pi}^*_{ij}$ is the $(i,j)$ entry of $\btil{\bF}^*$, we choose some $C \gtrsim 0$ and form $$\mathcal{S} = \{i: C<\btil{\pi}^*_{ij}<1-C, \forall j=1,...,n\}.$$ $\mathcal{S}$ identifies the rows of $\btil{\bF}^*$ where the $\logit$ function can be applied stably.  Here we use $C = \frac{1}{2n}$.
\item Define $\btil{\bF}_\mathcal{S}$ to be the corresponding subset of rows of $\btil{\bF}^*$, and calculate $\btil{\bL}_\mathcal{S} = \logit\left(\btil{\bF}_\mathcal{S} \right)$.  Let $\btil{\bL}_\mathcal{S}'$ be the row-wise mean centered and standard deviation scaled matrix $\btil{\bL}_\mathcal{S}$.
\item Perform SVD on $\btil{\bL}_\mathcal{S}'$ resulting in $\btil{\bL}_\mathcal{S}' = \bT \bLambda \bW^T$. Set $\bh{\bH}$ to be the $d \times n$ matrix composed of  the top $d-1$ right singular vectors of the SVD of $\bh{\bL}_\mathcal{S}'$ stacked on the row $n$-vector $(1, 1, \cdots, 1)$:
\begin{align*}
\bh{\bH} &= \begin{pmatrix}
\ & \ & \bW^{T}_{1:(d-1)} & \ & \ \\
1 & 1 & \cdots & 1 & 1 
\end{pmatrix}\\
&= \begin{pmatrix}
w_{11} & w_{21} & \cdots & w_{n1} \\
w_{12} & w_{22} & \cdots & w_{n2} \\
\vdots & \vdots & \  & \vdots \\
w_{1,d-1} & w_{2,d-1} & \cdots & w_{n,d-1} \\
1 & 1 & \cdots & 1
\end{pmatrix}.
\end{align*}
\end{enumerate}

\vspace{\baselineskip}
\noindent
\textbf{Algorithm 3}: Estimating $\bF$ and $\bL$ from LFA
\label{alg3}
\begin{enumerate}
\item Apply Algorithm 2 to $\bX$ to obtain $\bh{\bH}$.
\item For each SNP $i$, perform a logistic regression of the SNP genotypes $\bx_i = (x_{i1}, x_{i2}, \ldots, x_{in})$ on the rows of $\bh{\bH}$, specifically by maximizing the log-likelihood
$$
\ell(\bm{\pi}_i | \bx_i, \bh{\bH}) = \sum_{j=1}^n x_{ij} \log\left( \frac{\pi_{ij}}{1-\pi_{ij}} \right) + 2\log(1-\pi_{ij})
$$
under the constraint that $\logit(\pi_{ij}) = \sum_{k=1}^{d}  a_{ik} \bh{h}_{kj}$. It should be noted that an intercept is included because $\bh{h}_{dj}=1$ $\forall j$ by construction.
\item Set $\bh{a}_{ij}$ ($j=1, \ldots, n$) to be equal to the maximum likelihood estimates from the above model fit, for each of $i=1,\ldots, m$.  Let $\bh{\bL} = \bh{\bA} \bh{\bH}$, $\bh{\bF} = \logit^{-1}(\bh{\bL})$, and $\bh{\pi}_{ij}$ be the $(i,j)$ entry of $\bh{\bF}$:
$$
\bh{\pi}_{ij} = \frac{\exp \left\{ \sum_{k=1}^{d}  \bh{a}_{ik} \bh{h}_{kj} \right\}}{1 + \exp \left\{ \sum_{k=1}^{d}  \bh{a}_{ik} \bh{h}_{kj} \right\}}.
$$
\end{enumerate}

PCA-based estimation of Model 1 requires one application of singular value decomposition (SVD) and LFA requires two applications of SVD.  We leverage the fact that $n \gg d$ to utilize Lanczos bidiagonalization which is an iterative method for computing the singular value decomposition of a matrix \cn{Baglama2006}. Lanczos bidiagonalization excels at computing a few of the largest singular values and corresponding singular vectors of a sparse matrix. While the sparsity of genotype matrices is fairly low, we find that in practice using this method to perform the above estimation algorithms is more effective than using methods that require the calculation of all the singular values and vectors. This results in a dramatic reduction of the computational time needed for the implementation of our methods.

\section{{\sc Results}} 

We applied our methods to a comprehensive set of simulation studies and to the HGDP and TGP data sets.

\subsection{Simulation Studies} 

To directly evaluate the performance of the estimation methods (Section~\ref{algs}), we devised a simulation study where we generated synthetic genotype data with varying levels of complexity in population structure. Genotypes were simulated based on allele frequencies subject to structure from the BN model, the PSD model, spatially structure populations, and real data sets. For the first three types of simulations, the allele frequencies were parameterized by Model 1, while for the real data simulations, the allele frequencies were taken from model fits on the data themselves. 

A key property to assess is how well the estimation methods capture the overall structure. One way to evaluate this is to determine how well $\btil{\bS}$ from the PCA based method (Algorithm 1) estimates the true underlying $\bS$, and likewise how well $\bh{\bH}$ from LFA estimates the true $\bH$. Note that even though the genotype data was generated from the $\bF$ of Model 1, we can evaluate $\bh{\bH}$ by converting with $\bL = \logit(\bF)$. To evaluate PCA, we regressed each row of $\bF$ on $\btil{\bS}$ and calculated the average $R^2$; similarly, for LFA we regressed each row of $\bL$ on $\bh{\bH}$ and calculated the average $R^2$ value. The results are presented in Table \ref{tab:r2}. Both methods estimate the true latent structure well. 

\begin{figure}
\centerline{\includegraphics[width=0.85\textwidth]{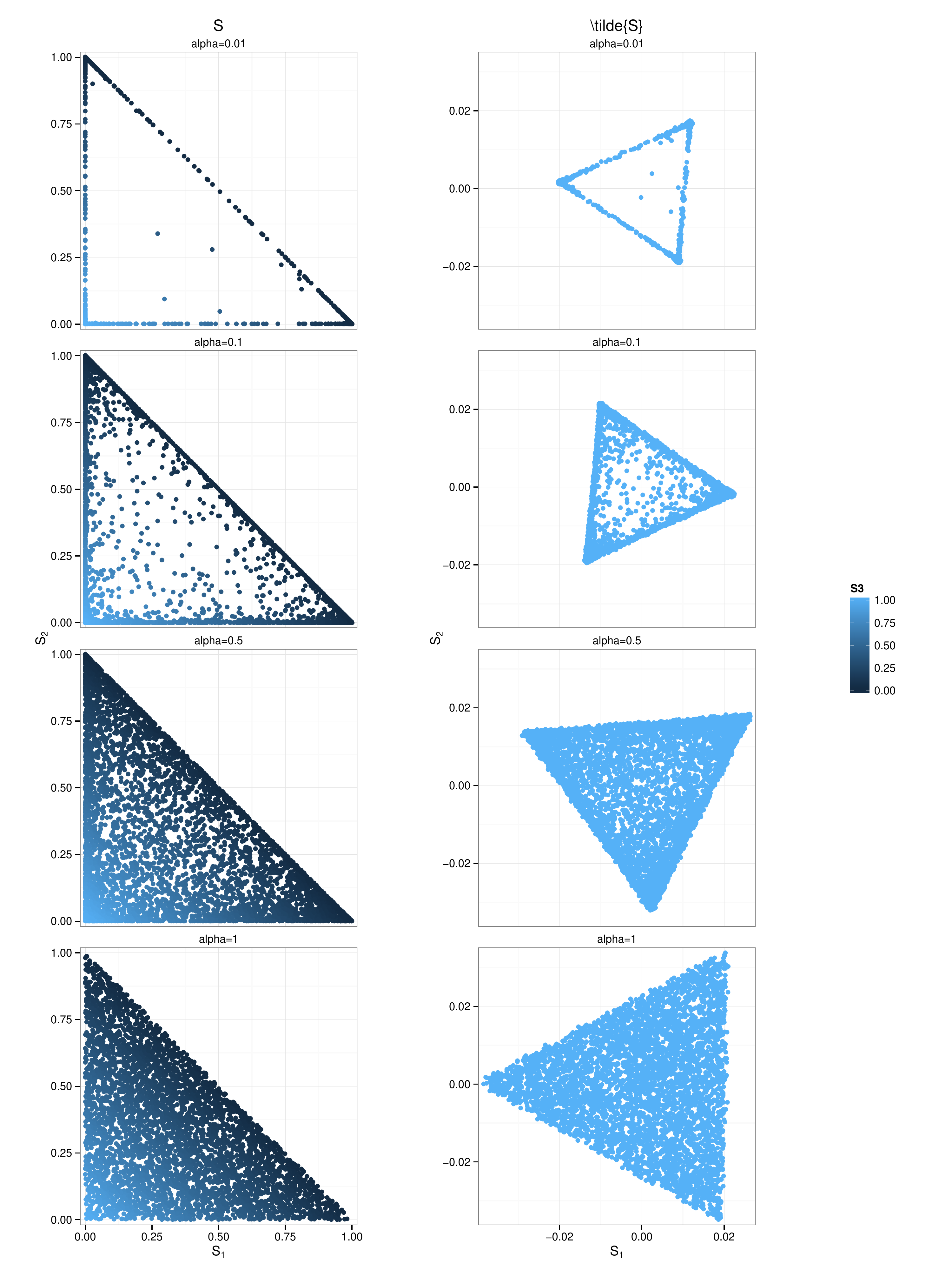}}
\caption{A mapping from $\bS$ to $\btil{\bS}$ for four simulated $\bS$ matrices under the PSD model.  The left column shows the simulated structure $\bS$ for each of four scenarios (a--d) and the right column shows the resulting estimated row basis of $\bS$ produced from PCA.  It can be seen that the scale on which $\bS$ was generated, all values in (0,1), is lost in the principal components, values in $\real$.}
\label{fig:PSDrange}
\end{figure}

We specifically note that when the PSD model was utilized to simulate structure, we were able to recover the structure $\bS$ very well (Supplementary Figure \ref{fig:PSDrange}) without needing to employ the computationally intensive and assumption-heavy Bayesian model fitting techniques from ref. \cn{PritchardStephens2000}.  Additionally, it seems that the $\btil{\bS}$ largely captures the geometry of $\bS$ where it may be the case that $\bS$ can be recovered with a high degree of accuracy by transforming $\btil{\bS}$ back into the simplex.  By comparing the results on the real data (Figures \ref{fig:HGDPpclf}-\ref{fig:TGPpclf}) with the simulated data (Supplementary Figure \ref{fig:PSDrange}), one is able to visually assess how closely the assumptions of the PSD model resemble real data sets.  When structure was simulated that differed substantially from the assumptions of the PSD model, our estimation methods were able to capture that structure just as well (Supplementary Figure \ref{fig:Spatialrange}).  This demonstrates the flexibility of the proposed approaches.  

We also compared PCA and LFA to two methods of fitting the PSD model, ADMIXTURE \cn{Alexander:2009p2792} and fastStructure \cn{Raj2013}, by seeing how well the methods estimated the individual specific allele frequencies $\pi_{ij}$ (Table \ref{tab:err}). For the real data scenarios, we generated synthetic genotypes based on estimates of $\bF$ from the four different methods, thus giving each method an opportunity to fit its own simulation. The methods were compared by computing three different error metrics with respect to the oracle $\bF$: Kullback-Leibler divergence, absolute error, and root mean squared error. PCA and LFA significantly outperformed ADMIXTURE and fastStructure, which confirms the intuitive understanding of the differences between the models: the goal of Model 1 and 2 is to estimate the allele frequencies $\piij$, while the PSD model provides a probabilistic interpretation of the structure by modeling them as admixture proportions.

The computational time required to perform the proposed methods was also significantly better than ADMIXTURE and fastStructure.  Both proposed methods completed calculations on average over 10 times faster than ADMIXTURE and fastStructure, with some scenarios as high as 150 times faster.  This is notable in that both ADMIXTURE and fastStructure are described as computationally efficient implementations of methods to estimate the PSD model \cn{Alexander:2009p2792,Raj2013}.

\subsection{Analysis of the HGDP and TGP Data}

\begin{figure}
\centerline{\includegraphics[width=\textwidth]{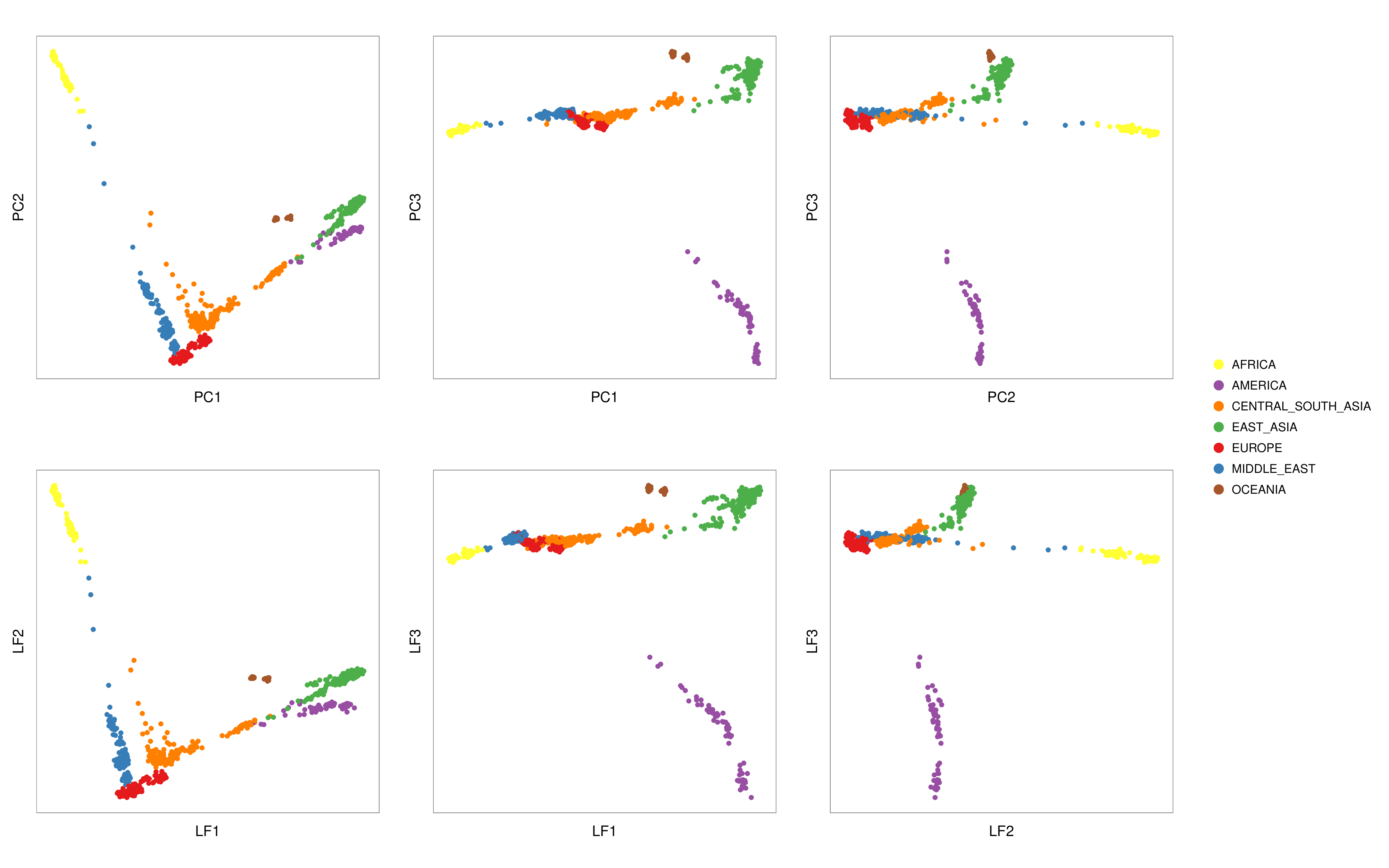}}
\caption{Principal components versus logistic factors for the HGDP data set.  The top three principal components from the HGDP data are plotted in a pairwise fashion in the top panel.  The top three logistic factors are plotted analogously in the bottom panel.  It can be seen that both approaches yield similar visualizations of structure.}
\label{fig:HGDPpclf}
\end{figure}

\begin{figure}
\centerline{\includegraphics[width=\textwidth]{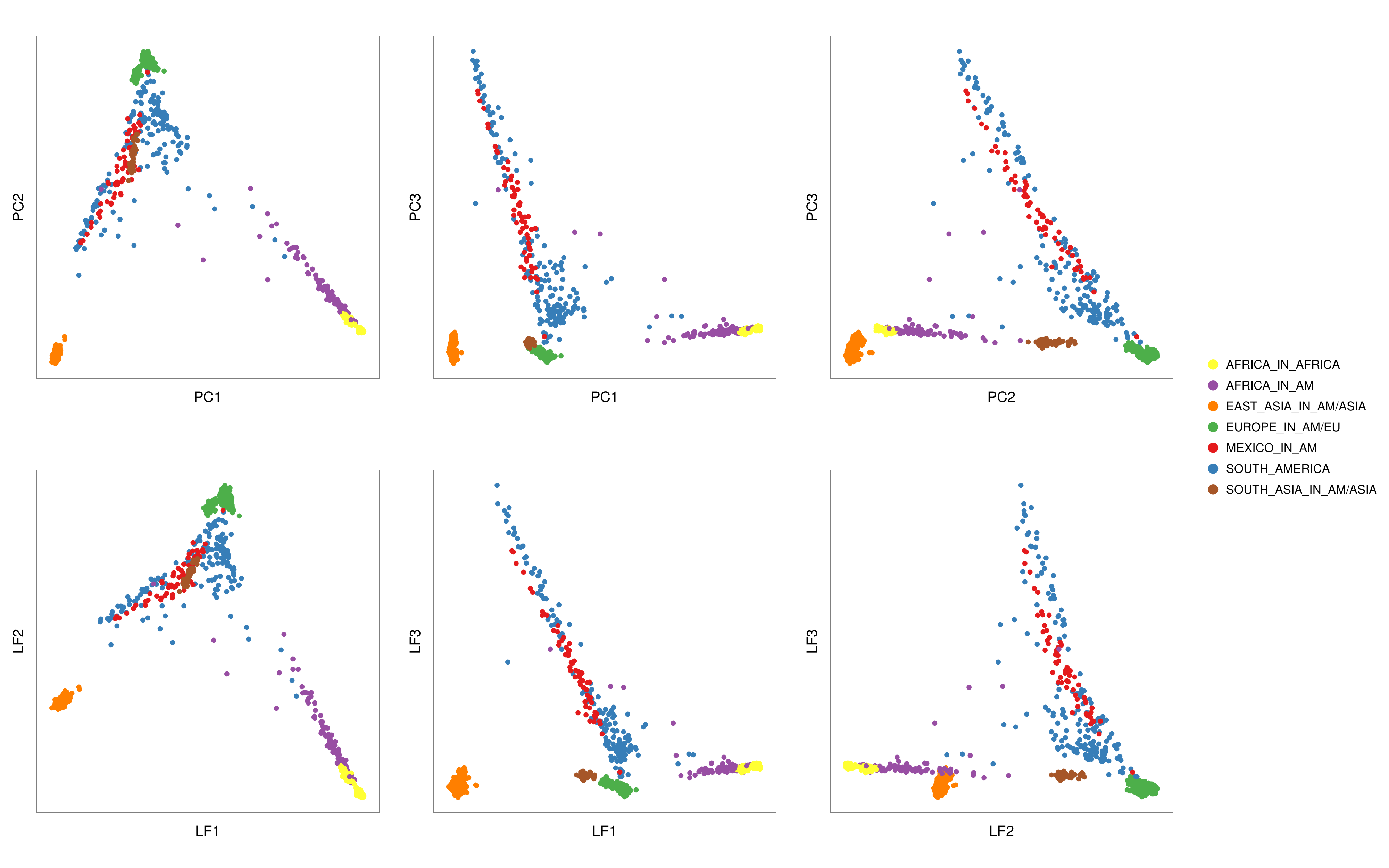}}
\caption{Principal components versus logistic factors for the TGP data set.  The top three principal components from the TGP data are plotted in a pairwise fashion in the top panel.  The top three logistic factors are plotted analogously in the bottom panel.  It can be seen that both approaches yield similar visualizations of structure.}
\label{fig:TGPpclf}
\end{figure}

We analyzed the HGDP and TGP data using the proposed methods.  The HGDP data consisted of $n=940$ individuals and $m=431,345$ SNPs, and the TGP data consisted of $n=1500$ and $m=339,100$ (see Supplementary Section~\ref{realdata} for details). We first applied PCA and LFA to these data sets and made bi-plots of the top three PCs and top three LFs (Figures \ref{fig:HGDPpclf} and \ref{fig:TGPpclf}).  It can be seen that PCA and LFA provide similar visualizations of the structure present in these data.  We next chose a dimension $d$ for the LFA model (Model 2) for each data set.  This was done by identifying the value of $d$ that provides the best overall goodness of fit with Hardy-Weinberg equilibrium (Supplementary Section~\ref{choosingd}).  We identified $d=15$ for HGDP and $d=7$ for TGP based on this criterion.  

One drawback of utilizing a PCA based approach (Algorithm 1) for estimating the individual-specific allele frequencies $\bF$ is that we are not guaranteed that all values of the estimates lie in $[0,1]$, so some form of truncation is necessary.  We found that 65.4\% of the SNPs in the HGDP data set and 26.5\% in the TGP data set resulted in at least one estimated individual-specific allele frequency $< 0$ or $> 1$ before the truncation was applied. Therefore, the truncation in forming the estimate $\widetilde{\bF}$ is necessary when employing Algorithm 1 to estimate $\bF$ from Model 1.  On the other hand, due to the formulation of Model 2, all estimated allele frequencies fall in the valid range when applying LFA (Algorithms 2 and 3).

The LFA framework provides a natural computational method for ranking SNPs according to how differentiated they are with respect to structure.  Note that existing methods typically require one to first assign each individual to one of $K$ discrete subpopulations \cn{Coop:2009p1530} which may make unnecessary assumptions on modern data sets such as HGDP and TGP.  In order to rank SNPs for differentiation, we calculate the deviance statistic when performing a logistic regression of the SNPs genotypes on the logistic factors.  Specifically we calculated the deviance by comparing the models $\logit(\bm{\pi}_i) = a_{id} \bm{h}_d$ vs. $\logit(\bm{\pi}_i) = \sum_{k=1}^{d} a_{ik} \bm{h}_k$, where the former model is intercept only (i.e., $d=1$, no structure).   

Our application of LFA to identify SNPs with allele frequencies differentiated according to structure can be developed further. First, the recently proposed ``jackstraw'' approach \cn{Chung2013} provides a manner in which statistical significance can be assigned to these SNPs.  Assigning statistical significance to the population differentiation of SNPs has traditionally been a difficult problem \cn{Akey2002}. Second, we found the deviance measure tends to have more extreme values for SNPs with larger minor allele frequencies (MAFs).  Therefore, the ranking of SNPs may be made more informative if MAF is taken into account.  Third, although this ranking is identifying differentiation and not specifically selection, it may provide a useful starting point in understanding methods that attempt to detect selection.

\begin{sidewaysfigure}
\centerline{\includegraphics[width=\textwidth]{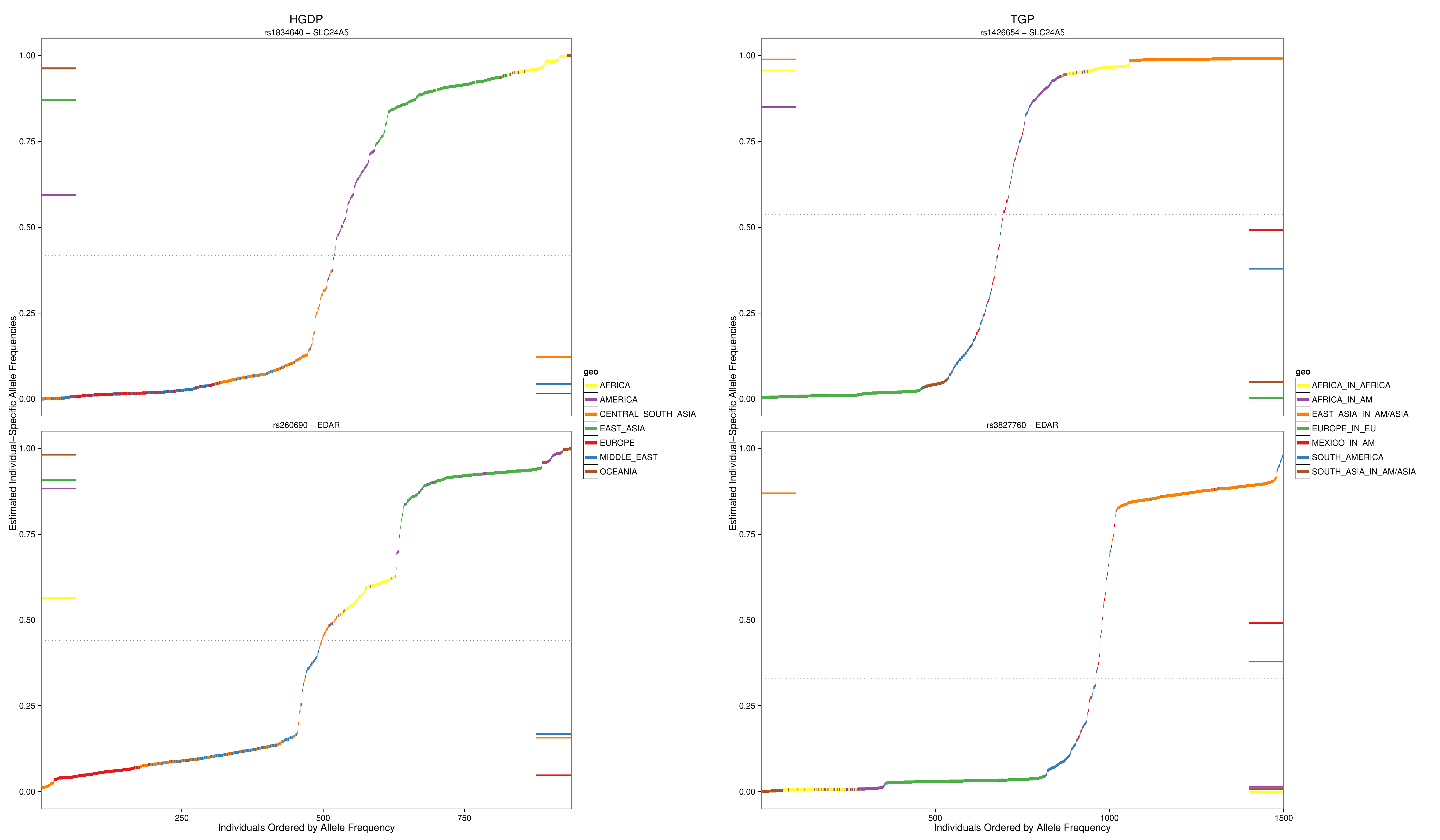}}
\caption{SNPs with highly differentiated allele frequencies with respect to structure.  Two of the most highly different SNPs according to LFA are shown for the HGDP and TGP data sets.  For each SNP, the $\bh{\pi}_{ij}$ values are ordered and they are colored according reported ancestry.  The horizontal bars on the sides of the plots denote the usual allele frequency estimates formed within each ancestral group.}
\label{fig:topSNPs}
\end{sidewaysfigure}

The most differentiated SNPs (Supplementary Tables \ref{tab:HGDPtop} and \ref{tab:TGPtop}) reveal some noteworthy results, especially considering the flexible approach to forming the ranking.  SNPs located within or very close to {\em SLC24A5} were the top ranked in both HGDP and TGP.  This gene is well known to be involved in determining skin pigmentation in humans \cn{Lamason:2005fj} and is hypothesized to have been subject to positive selection \cn{Sabeti2007}.  The next most highly ranked SNPs in both studies are located in {\em EDAR}, which plays a major role in distinguishing phenotypes (e.g., hair follicles) among Asians.  SNP rs3827760 is the second most differentiated SNP in the TGP data, which has also been hypothesized to be under positive selection in humans and whose causal role in the hair follicle phenotype has been verified in a mouse model \cn{Kamberov:2013p2731}.  SNPs corresponding to these two genes for both studies are plotted in increasing order of $\bh{\pi}_{ij}$ values, revealing subtle variation within each major ancestral group in addition to coarser differences in allele frequency (Figure~\ref{fig:topSNPs}).  Other noteworthy genes with highly differentiated proximal SNPs include: 
\begin{itemize}
\item {\em FOXP1}, which is a candidate gene for involvement in tumor progression and plays an important regulatory role with {\em FOXP2} \cn{Banham2001,Shigekawa2011};
\item {\em TBC1D1} in which genetic variation has been shown to confer risk for severe obesity in females \cn{Stone2006};
\item {\em KIF3C}, a novel kinesin-like protein, which has been hypothesized to be involved in microtubule-based transport in neuronal cells \cn{Sardella1998};
\item {\em KCNMA1}, a recently identified susceptibility locus for obesity \cn{Jiao2011};
\item {\em CTNNA3} in which genetic variation has been shown to be associated with diisocyanate-induced occupational asthma \cn{Bernstein2013}; 
\item {\em PTK6}, breast tumor kinase (Brk), which is known to function in cell-type and context-dependent processes governing normal differentiation \cn{Ostrander2010}. 
\end{itemize}
We have provided information on the 5000 most differentiated SNPs for both TGP and HGDP in supplementary files.

\subsection{Software}

An R package called \texttt{lfa} is available at \url{https://github.com/StoreyLab/lfa}.

\section{{\sc Discussion}}  

We have investigated two latent variable models of population structure to simultaneously estimate all individual-specific allele frequencies from genome-wide genotyping data.  Model 1, a direct model of allele frequencies, can be estimated by using a modified PCA and Model 2, a model of the $\logit$ transformation of allele frequencies, is estimated through a new approach we called ``logistic factor analysis'' (LFA).  For both models, the latent variables are estimated in a nonparametric fashion, meaning we do not make any assumptions about the underlying structure captured by the latent variables.  These models are general in that they allow for each individual's genotype to be generated from an allele frequency specific to that individual, which includes discretely structured populations, admixed populations, and spatially structured populations.  In LFA, we construct a model of the $\logit$ of these allele frequencies in terms of underlying factors that capture the population structure.  We have proposed a computationally efficient method to estimate this model that requires only two applications of SVD.  This approach builds on the success of PCA in that we are able to capture population structure in terms of a low-dimensional basis.  It improves on PCA in that the latent variables we estimate can be straightforwardly incorporated into downstream statistical inference procedures that require well-behaved estimates of allele frequencies.  In particular, statistical inferences of Hardy-Weinberg equilibrium, $\fst$, and marker-trait associations are amenable to complex population structures within our framework. 

We demonstrated our proposed approach on the HGDP and TGP data sets and several simulated data sets motivated by the HapMap, HGDP, and TGP data sets as well as the PSD model and spatially distributed structures.  It was shown that our method estimates the underlying logistic factors with a high degree of accuracy.  We also showed that applying PCA to genotype data estimates a row basis of population structure on the original allele frequency scale to a high degree of accuracy.  However, problems occur when trying to recover estimates of individual-specific allele frequencies because PCA is a real-valued model that does not always result in allele frequency estimates lying between 0 and 1.  

Although PCA has become very popular for genome-wide genotype data, it should be stressed that PCA is fundamentally a method for characterizing variance and special care should be taken when applying it to estimate latent variables.  The authoritative treatment of PCA \cn{Jolliffe10} eloquently makes this point throughout the text and considers cases where factor analysis is more appropriate than PCA through examples reminiscent of the population structure problem.  Here, we have shown that modeling and estimating population structure can be understood from the factor analysis perspective, leading to estimates of individual-specific allele frequencies through their natural parameter on the $\logit$ scale.  At the same time, we have avoided some of the difficulties of traditional parameteric factor analysis by maintaining the relevant nonparametric properties of PCA, specifically in making no assumptions about the underlying probability distributions of the logistic factors that capture population structure.

\clearpage
\section{{\sc Figures and Tables}} 

\clearpage

\clearpage

\clearpage

\clearpage
\begin{table}
\caption{Accuracy in estimating linear bases for $\bS$.  Column 1 shows the scenario from which the data were simulated. Columns 2 and 3 display the estimation accuracy of the PCA based method (Column 2) and LFA (Column 3).  Column 2 shows the mean $R^2$ value when regressing the true $(\pi_{i1}, \pi_{i2}, \ldots, \pi_{in})$ on $\bh{\bS}$ from PCA, averaging across all SNPs. Column 3 shows the mean $R^2$ value when regressing the true $\left(\logit(\pi_{i1}), \logit(\pi_{i2}), \ldots, \logit(\pi_{in})\right)$ on $\bh{\bH}$ from LFA, averaging across all SNPs. All estimated standard errors fell between $10^{-6}$ and $10^{-8}$ so these are not shown. Note for each scenario, $R^2$ values are higher for the method from which the true $\bF$ matrix was generated. All but the two scenarios marked with an asterisk (*) are from Model 1, while the two marked scenarios are from Model 2, where we took $\bF = \logit ^{-1} \bL$. }
\label{tab:r2}

\ 
\begin{center}
\begin{tabular}{|l|cc|cc|}\hline
&\multicolumn{2}{c|}{Mean $R^2$} \\
\hline
\rule{0pt}{3ex} Scenario & $\bF \sim \btil{\bS}$ & $\logit(\bF) \sim \bh{\bH}$ \\ 
\hline\hline
TGP fit by PCA    & 0.9998 & 0.9722 \\
TGP fit by LFA *  & 0.9912 & 0.9990 \\
HGDP fit by PCA   & 0.9996 & 0.9614 \\
HGDP fit by LFA * & 0.9835 & 0.9983 \\
BN                & 0.9999 & 0.9999 \\
PSD $\alpha=0.01$ & 0.9998 & 0.9974 \\
PSD $\alpha=0.1$  & 0.9998 & 0.9879 \\
PSD $\alpha=0.5$  & 0.9996 & 0.9827 \\
PSD $\alpha=1$    & 0.9993 & 0.9844 \\
Spatial $a=0.1$   & 0.9999 & 0.9964 \\
Spatial $a=0.25$  & 0.9999 & 0.9962 \\
Spatial $a=0.5$   & 0.9999 & 0.9964 \\
Spatial $a=1$     & 0.9998 & 0.9970 \\
\hline
 \end{tabular}
\end{center}
\end{table}

\begin{sidewaystable}
\caption{Accuracy in estimating $\pi_{ij}$ parameters by the PCA based method and LFA. Each row is a different simulation scenario. Each column is the accuracy of a method's fits with the given metric.} 
\label{tab:err}
\begin{tabular*}{\linewidth}{ @{\extracolsep{\fill}} ll |cccc|cccc|cccc @{}}
\toprule
\multicolumn{2}{c}{Scenario} & \multicolumn{4}{c}{Median KL} &
\multicolumn{4}{c}{Mean Abs. Err.} & \multicolumn{4}{c}{RMSE}  \\
\midrule \midrule \addlinespace \\
 & & PCA & LFA & ADX & FS & PCA & LFA & ADX & FS & PCA & LFA & ADX & FS \\
\cmidrule{3-6} \cmidrule{7-10} \cmidrule{11-14}
\addlinespace \\
\multirow{1}{*}{\rotatebox{90}{BN}} 
&  & 6.9\e{-5} & 6.8\e{-5} & 2.6\e{-3} & 2.6\e{-3}   & 5.8\e{-3} & 5.8\e{-3} & 3.7\e{-2} & 3.7\e{-2}   & 7.5\e{-3} & 7.5\e{-3} & 5.8\e{-2} & 5.8\e{-2} \\
\midrule \addlinespace \\
\multirow{4}{*}{\rotatebox{90}{PSD}} 
& $\alpha=0.01$  & 7.0\e{-5} & 7.3\e{-5} & 1.6\e{-2} & 1.6\e{-2}   & 5.6\e{-3} & 5.8\e{-3} & 9.7\e{-2} & 9.7\e{-2}   & 7.2\e{-3} & 7.6\e{-3} & 1.7\e{-1} & 1.7\e{-1} \\
& $\alpha=0.1$   & 6.7\e{-5} & 9.2\e{-5} & 3.6\e{-2} & 3.6\e{-2}   & 5.6\e{-3} & 6.9\e{-3} & 1.6\e{-1} & 1.6\e{-1}   & 7.2\e{-3} & 9.3\e{-3} & 2.4\e{-1} & 2.4\e{-1} \\
& $\alpha=0.5$   & 6.3\e{-5} & 8.5\e{-5} & 5.4\e{-2} & 5.4\e{-2}   & 5.6\e{-3} & 6.8\e{-3} & 1.4\e{-1} & 1.4\e{-1}   & 7.3\e{-3} & 9.0\e{-3} & 1.8\e{-1} & 1.8\e{-1} \\
& $\alpha=1.0$   & 6.1\e{-5} & 7.4\e{-5} & 3.3\e{-2} & 3.3\e{-2}   & 5.6\e{-3} & 6.3\e{-3} & 1.4\e{-1} & 1.4\e{-1}   & 7.4\e{-3} & 8.4\e{-3} & 2.2\e{-1} & 2.2\e{-1} \\
\midrule \addlinespace \\
\multirow{4}{*}{\rotatebox{90}{Spatial}} 
& $a=0.1$   & 7.3\e{-5} & 1.2\e{-4} & 8.2\e{-3} & 8.1\e{-3}   & 5.5\e{-3} & 7.6\e{-3} & 7.4\e{-2} & 7.4\e{-2}   & 7.0\e{-3} & 1.0\e{-2} & 1.2\e{-1} & 1.2\e{-1} \\
& $a=0.25$  & 6.9\e{-5} & 1.1\e{-4} & 8.6\e{-3} & 8.6\e{-3}   & 5.6\e{-3} & 7.4\e{-3} & 9.3\e{-2} & 9.3\e{-2}   & 7.2\e{-3} & 9.8\e{-3} & 1.6\e{-1} & 1.6\e{-1} \\
& $a=0.5$   & 6.6\e{-5} & 9.5\e{-5} & 1.0\e{-2} & 1.0\e{-2}   & 5.6\e{-3} & 6.9\e{-3} & 6.7\e{-2} & 6.7\e{-2}   & 7.2\e{-3} & 9.2\e{-3} & 1.0\e{-1} & 1.0\e{-1} \\
& $a=1.0$   & 6.3\e{-5} & 7.8\e{-5} & 1.2\e{-2} & 1.2\e{-2}   & 5.7\e{-3} & 6.4\e{-3} & 1.1\e{-1} & 1.1\e{-1}   & 7.4\e{-3} & 8.5\e{-3} & 1.7\e{-1} & 1.7\e{-1} \\
\midrule \addlinespace \\
\multirow{4}{*}{\rotatebox{90}{TGP fit}} 
& PCA     & 4.1\e{-4} & 5.2\e{-4} & 2.8\e{-3} & 3.4\e{-3}   & 1.3\e{-2} & 1.5\e{-2} & 8.1\e{-2} & 8.3\e{-2}   & 1.8\e{-2} & 2.1\e{-2} & 1.5\e{-1} & 1.5\e{-1} \\
& LFA     & 4.3\e{-4} & 4.8\e{-4} & 2.4\e{-3} & 2.7\e{-3}   & 1.3\e{-2} & 1.4\e{-2} & 7.9\e{-2} & 8.1\e{-2}   & 1.8\e{-2} & 2.0\e{-2} & 1.4\e{-1} & 1.5\e{-1} \\
& ADX     & 5.4\e{-4} & 4.4\e{-4} & 5.0\e{-3} & 5.5\e{-3}   & 1.5\e{-2} & 1.3\e{-2} & 1.1\e{-1} & 1.1\e{-1}   & 2.0\e{-2} & 1.9\e{-2} & 2.0\e{-1} & 2.0\e{-1} \\
& FS      & 4.1\e{-4} & 5.5\e{-4} & 7.8\e{-4} & 9.2\e{-4}   & 1.3\e{-2} & 1.5\e{-2} & 5.6\e{-2} & 5.8\e{-2}   & 1.8\e{-2} & 2.1\e{-2} & 1.3\e{-1} & 1.3\e{-1} \\
\midrule \addlinespace \\
\multirow{4}{*}{\rotatebox{90}{HGDP fit}} 
& PCA     & 1.0\e{-3} & 1.2\e{-3} & 1.3\e{-2} & 1.4\e{-2}   & 2.3\e{-2} & 2.5\e{-2} & 1.2\e{-1} & 1.2\e{-1}   & 3.4\e{-2} & 3.6\e{-2} & 2.2\e{-1} & 2.2\e{-1} \\
& LFA     & 9.9\e{-4} & 1.1\e{-3} & 1.3\e{-2} & 1.2\e{-2}   & 2.2\e{-2} & 2.4\e{-2} & 1.2\e{-1} & 1.2\e{-1}   & 3.5\e{-2} & 3.7\e{-2} & 2.2\e{-1} & 2.2\e{-1} \\
& ADX     & 1.6\e{-3} & 1.4\e{-3} & 2.3\e{-3} & 2.3\e{-3}   & 2.6\e{-2} & 2.6\e{-2} & 5.6\e{-2} & 5.6\e{-2}   & 3.6\e{-2} & 3.7\e{-2} & 1.0\e{-1} & 1.0\e{-1} \\
& FS      & 1.4\e{-3} & 1.6\e{-3} & 3.1\e{-2} & 2.9\e{-2}   & 2.6\e{-2} & 2.7\e{-2} & 1.4\e{-1} & 1.3\e{-1}   & 3.6\e{-2} & 3.8\e{-2} & 2.2\e{-1} & 2.1\e{-1} \\
\bottomrule
\end{tabular*}
\end{sidewaystable}

\clearpage

\section{{\sc Supplementary Material}}  
\subsection{Data sets}
\label{realdata}
The HGDP data set was constructed by intersecting the data available from the HGDP web site, \url{http://www.hagsc.org/hgdp/files.html}, with the set of individuals ``H952'' identified by Rosenberg (2006) \cn{Rosenberg2006hgdp} with a high confidence as containing no first and second-degree relative pairs.  This yielded complete SNP genotype data on 431,345 SNPs for 940 individuals.

In order to obtain data from the TGP we first obtained the genotype data that had been measured through the Omni Platform, 2011-11-17, \url{ftp://ftp.1000genomes.ebi.ac.uk/vol1/ftp/technical/working}. 
We removed related individuals based on the TGP sample information. 
We then sorted individuals according to least percentage of SNPs with missing data, and we selected the top 1500 individuals.  This yielded complete SNP genotype data on 339,100 SNPs for 1500 individuals.

We utilized the HapMap data set in the simulated data described below.  We obtained the HapMap data release 23a, NCBI build 36 from \url{www.hapmap.org} consisting of unrelated individuals: 60 from European ancestry group (CEU), 60 from Yoruba, Africa (YRI) , and 90 from Japan and China (JPT+CHB).  We identified all SNPs with observed minor allele frequency $\geq 5\%$ and with no missing data. The total number of SNPs used after filtering in each population were CEU: 1,416,940, YRI: 1,539,314, JPT+CHB: 759,452. We then identified all SNPs common to all three populations resulting in a total of 363,955.

\subsection{Choosing the model dimension}
\label{choosingd}
The model dimension $d$ was determined for the HGDP and TGP data sets under the rationale that when $d$ is large enough, then the great majority of SNPs should appear to be in HWE.  When $d$ is too small, then the structure which has not been accounted for will lead to spurious deviations from HWE.  Values $d=1, 2, \ldots, 20$ were considered for each data set, and we ended up identifying $d=15$ for HGDP and $d=7$ for TGP.  We note that these choices could also be interpreted as reasonable according to a scree plot when PCA was applied to the genotype data.  

For a given $d$ value, we formed $\bh{\bF}$ using the LFA method.  We calculated a HWE goodness of fit statistic for each SNP $i$ as follows:
$$
\sum_{k=0}^2 \frac{\left[ \sum_{j=1}^n 1(x_{ij}=k) - \sum_{j=1}^n  {2 \choose k} \bh{\pi}_{ij}^k (1-\bh{\pi}_{ij})^{2-k} \right]^2}{\sum_{j=1}^n  {2 \choose k} \bh{\pi}_{ij}^k (1-\bh{\pi}_{ij})^{2-k}},
$$
where $\sum_{j=1}^n 1(x_{ij}=k)$ is the observed number of genotypes equal to $k$ and $\sum_{j=1}^n {2 \choose k} \bh{\pi}_{ij}^k (1-\bh{\pi}_{ij})^{2-k}$ is the expected number of genotypes equal to $k$ under HWE.  We then utilized $\bh{\bF}$ to simulate five instances of a genotype matrix $\bX^0$ under HWE where we simulated $x^0_{ij} \sim \mbox{Binomial}(2, \bh{\pi}_{ij})$.  On each simulated genotype matrix $\bX^0$, we again applied LFA to obtain $\bh{\bF}^0$ and calculate HWE goodness of fit statistics.  These goodness of fit statistics were then pooled across all five simulated data sets and across all SNPs to form the null distribution, which then allowed us to calculate a HWE p-value for each observed SNP.  (It should be noted that we also formed a separate null distribution according to minor allele frequency bins of length 0.05, and we arrived at the same conclusion.)  We then compared these p-values to the Uniform(0,1) distribution and also against the p-values from the $d+1$ case. This allowed us to identify a value of $d$ where the HWE p-values were both close to the Uniform(0,1) distribution and to the HWE p-values from the $d+1$ case.   

\subsection{Simulated data}

For each simulation scenario, genotypes $\bX$ were simulated such that $x_{ij} \sim \mbox{Binomial}(2, \pi_{ij})$, where $\pi_{ij}$ were elements of the allele frequency matrix $\bF$. The results from the simulated data are summarized in Tables 1 and 2. 

\subsub{Balding-Nichols (BN).} For each SNP in the HapMap data set, we estimated its marginal allele frequency according to the observed frequency and estimated its $\fst$ value using the Weir \& Cockerham estimate \cn{weir84}.  We set the simulated data to have $m=100,000$ SNPs and $n=5000$ individuals with $d=3$. Using Model 1, the $\bS$ matrix was generated by sampling its columns $\bs^j$ i.i.d. from $(1,0,0)^T$, $(0,1,0)^T$, and $(0,0,1)^T$ with respective probabilities $60/210$, $60/210$, and $90/210$ to reflect the original data's subpopulation proportions.  For each row $i$ of $\bG$, we simulated i.i.d. draws from the Balding-Nichols model:  $\bgamma_{i1}, \bgamma_{i2}, \bgamma_{i3} \iid \mbox{BN}(p_i, F_i)$, where the pair $(p_i, F_i)$ was randomly selected from among the marginal allele frequency and $\fst$ pairs calculated on the HapMap data set.  

\subsub{PSD.} We analyzed each SNP in the HGDP data set to estimate its marginal allele frequency according to the observed marginal frequency and $\fst$ using the Weir \& Cockerham estimate \cn{weir84}.  To estimate $\fst$, each individual in the HGDP data set was assigned to one of $K=5$ subpopulations according to the analysis in Rosenberg et al. (2002) \cn{Rosenberg2002}.  We set $m=100,000$ SNPs and $n=5000$ individuals with $d=3$.  Again utilizing Model 1, each row $i$ of $\bG$ was simulated according to $\gamma_{i1}, \gamma_{i2}, \gamma_{i3} \iid \mbox{BN}(p_i, F_i)$, where the pair $(p_i, F_i)$ was randomly selected from among the marginal allele frequency and $\fst$ pairs calculated on the HGDP data set.  To generate $\bS$, we simulated $(s_{1j}, s_{2j}, s_{3j}) \iid \mbox{Dirichlet}(\bmm{\alpha})$ for $j=1,\ldots,5000$.  We considered $\bmm{\alpha} = (0.01, 0.01, 0.01)$, $\bmm{\alpha} = (0.1, 0.1, 0.1)$, $\bmm{\alpha} = (0.5, 0.5, 0.5)$, and $\bmm{\alpha} = (1, 1, 1)$. It should be noted that as $\bmm{\alpha} \rightarrow \bmm{0}$, the draws from the Dirichlet distribution become increasingly closer to assigning each individual to one of three discrete subpopulations with equal probability.  When $\bmm{\alpha} = (1, 1, 1)$, the admixture proportions are distributed uniformly over the simplex.

\begin{figure}
\centerline{\includegraphics[width=0.85\textwidth]{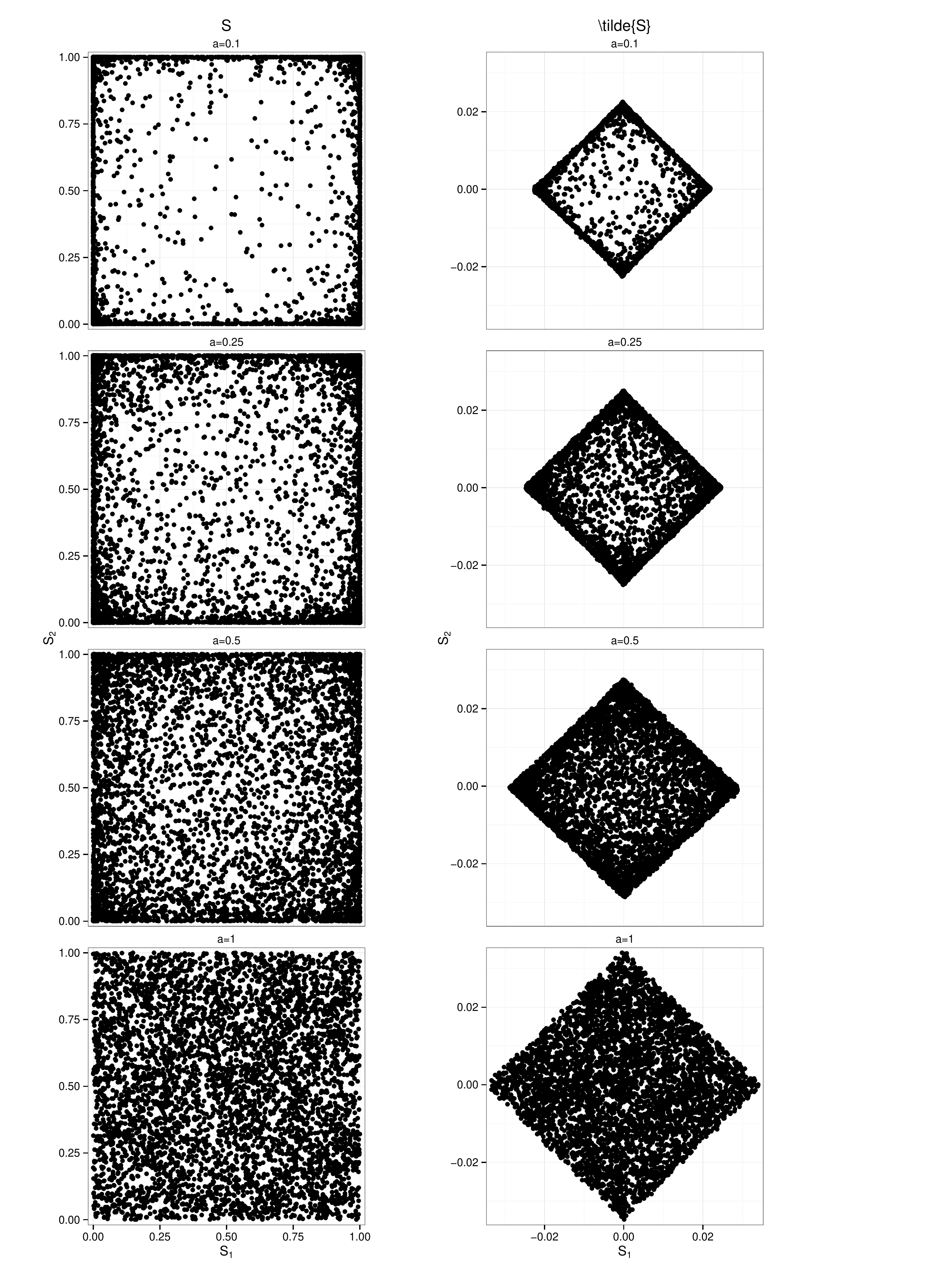}}
\caption{A mapping from $\bS$ to $\btil{\bS}$ for four simulated $\bS$ matrices under the Spatial model.  The left column shows the simulated structure $\bS$ for each of four scenarios (a--d) and the right column shows the resulting estimated row basis of $\bS$ produced from PCA.  It can be seen that the scale on which $\bS$ was generated, all values in (0,1), is lost in the principal components, values in $\real$.}
\label{fig:Spatialrange}
\end{figure}

\subsub{Spatial.} This scenario is meant to create population structure that is driven by spatial position of the individual.  We set the simulated data to have $m=100,000$ SNPs and $n=5000$ individuals with $d=3$.  Rows $i=1,2$ of $\bS$ were simulated as $s_{ij} \iid \mbox{Beta}(a, a)$ for $j=1, \ldots, 5000$, and row 3 of $\bS$ contained the intercept term, $s_{3j} = 1$.  We considered four values of $a$:  0.1, 0.25, 0.5, and 1.  The first two rows of $\bS$ place each individual in a two-dimensional space (Figure \ref{fig:Spatialrange}), where the ancestry of individual $j$ is located at $(s_{1j}, s_{2j})$ in the unit square.  When $a=1$, the Beta$(a,a)$ distribution is Uniform$(0,1)$, so this scenario represents a uniform distribution of individuals in unit square.  As $a \rightarrow 0$, the Beta$(a,a)$ places each individual with equal probabilities in one of the four corners of the unit square.  The matrix $\bG$ was created by sampling $\gamma_{ij} \iid 0.9 \times \mbox{Uniform}(0, 1/2)$ for $j=1,2$ and $\gamma_{i3} = 0.05$.  It should be noted that all $\pi_{ij} \in [0.05, 0.95]$ by construction.

\subsub{Real Data.} For the HGDP and TGP scenarios, we estimated an allele frequency matrix $\bF$ from the real data via four different methods. For HGDP we had $m=431,345$ SNPs by $n=940$ individuals with $d=15$, and for TGP we had $m=339,100$ and $n=1,500$ with $d=7$. The four methods are: 

\begin{itemize}
\item {\em PCA}: $\bF$ was taken to be the matrix $\btil{\bF}$ estimated via Algorithm 1.
\item {\em LFA}: $\bF = \logit^{-1}(\bh{\bL})$, where $\bh{\bL}$ was estimated via Algorithm 3.
\item {\em ADX}: $\bF$ was taken to be the matrix formed by computing the marginal allele frequencies in the Pritchard-Stephens-Donnelly model, i.e. $\bF = \bP \bQ$, and $\bP$ and $\bQ$ were estimated via the software ADMIXTURE \cn{Alexander:2009p2792}.
\item {\em FS}: Same as above except $\bP$ and $\bQ$ are estimated via the software fastStructure \cn{Raj2013}.
\end{itemize}

\subsection{Error Measures Used to Evaluate Estimates of $\bF$ and $\bL$}
Estimates of $\pi_{ij}$ were evaluated with three different metrics.  Let $\bh{\pi}_{ij}$ be the estimate for any given method.

The {\em Kullback-Leibler divergence} for the binomial distribution allows us to measure the difference between the distribution from the estimated allele frequencies to the distribution from the oracle allele frequencies:

$$
\mbox{KL} = \piij \ln\left( \frac{\piij}{\bh{\pi}_{ij}}\right) + (1-\piij) \ln\left( \frac{1-\piij}{1-\bh{\pi}_{ij}}\right).
$$

\noindent {\em Mean absolute error} compares the allele frequencies directly:

$$
\mbox{MAE} = \frac{1}{m \times n} \sum_{i=1}^{m} \sum_{j=1}^{n} \left| \pi_{ij} - \bh{\pi}_{ij} \right|.
$$

\noindent {\em Root mean squared error}:

$$
\mbox{RMSE} = \sqrt{\frac{1}{m \times n} \sum_{i=1}^{m} \sum_{j=1}^{n} \left( \logit(\pi_{ij}) - \logit(\bh{\pi}_{ij}) \right)^2}.
$$

\subsection{$\fst$ for individual-specific allele frequencies} 
\label{fst}
By considering the derivation of $\fst$ for $K$ discrete populations as described in Weir (1984, 1996) \cn{Weir1996,weir84}, it can be seen that a potential generalization of $\fst$ to arbitrary population structure is 
$$
\fst = 1 - \frac{\e_{\bZ}[\var(x | \bZ)]}{\var(x)},
$$
where, as described in Section \label{allelefreqs}, $\bZ$ is a latent variable capturing an individual's population structure position or membership.  The allele frequency of a SNP conditional on $\bZ$ can be viewed as being a function of $\bZ$, which we have denoted by $\pi(\bZ)$.  If $n$ individuals are sampled independently and homogeneously from the population\footnote{When the individuals are not sampled homogeneously throughout the population (e.g., in the HapMap data with 60, 60, and 90 observations from three discretely defined subpopulations), then it may be the case that the above quantity should be modified to reflect the stratified or non-homogeneous sampling.} such that $\bz_1, \ldots, \bz_n$ are i.i.d. from the distribution on $\bZ$, then for SNP $i$ in HWE, it follows that $\var(x_{ij} | z_j) = 2 \pi_{ij} (1-\pi_{ij})$ and 
$$
\fst \stackrel{a.s.}{=} \lim_{n \rightarrow \infty} 1-\frac{\frac{1}{n}\sum_{j=1}^n \pi_{ij} (1-\pi_{ij})}{\overline{\pi}_i (1-\overline{\pi}_i)},
$$
where $\overline{\pi}_i = \sum_{j=1}^n \pi_{ij}/n$ is the marginal allele frequency among the $n$ individuals.  Thus, good estimates of the $\pi_{ij}$ values may be useful for estimating $\fst$ in this general setting. One example would be to form a plug-in estimate of $\fst$ by replacing $\pi_{ij}$ with $\bh{\pi}_{ij}$ from the proposed LFA method.

\subsection{Relationship of LFA to existing models and methods} 
\label{existingmethods}
The problem of modeling a genotype matrix $\bX$ in order to uncover latent variables that explain cryptic structure is a special case of a much more general problem that has been considered for several years in the statistics literature \cn{Bartholomew1984,Moustaki2000}. Under a latent variable model, it is assumed that the ``manifest'' (observed) variables are the result of the ``latent''  (unobserved) variables. Different types of the latent variable models can be grouped according to whether the manifest and latent variables are categorical or continuous. For example, factor analysis is a latent variable method for the case where both manifest variable and latent variable are continuous.  A proposed naming convention \cn{BKM2011} is summarized as follows:

\begin{center}
\begin{tabular}{|c|cc|}
\hline
\ & \multicolumn{2}{c|}{Manifest variables} \\
Latent variables & Continuous & Categorical \\
\hline
Continuous &Factor analysis & Latent trait analysis \\
Categorical & Latent profile analysis & Latent class analysis \\
\hline
\end{tabular}
\end{center}
The problem we consider is that the manifest variables (observed gentoypes) are categorical, and they are driven by latent variables (population structure) that may either be categorical (discrete population structure) or continuous (complex population structure).  Therefore, the LFA method may be described as a nonparametric latent variable estimation method that jointly captures latent trait analysis and latent class analysis.  Another naming convention that we could apply to LFA would be to call it a nonparametric latent variable model for Binomial data.  The naming conventions of latent variable models are inconsistent and often confusing \cn{BKM2011}.

Bartholomew (1980) \cn{Batholomew1980} proposed a model related to equation (2) to identify latent variables that influence the probabilities of a collection of Binomial random variables.  See also Bartholomew et al. 2011 for a comprehensive treatment of this area, which they call ``general linear latent variable models'' (GLLVM). In particular, when the manifest variables $x_{ij} \sim$ Bernoulli($\pi_{ij}$) and the latent variables $h_{kj}$ are continuous variables, the GLLVM in this case is Model 2 , $\logit(\pi_{ij}) = \sum_{k=1}^d a_{ik} h_{kj}$. While we begin with this model, there are some key differences.  The number of manifest variables in the data considered in Bartholomew (1980) and related work is notably smaller than genome-wide genotype data, so the assumptions and estimation approach differ substantially.  Model assumptions are typically made about the probability distributions of the latent variables; we consider these model assumptions too strong and also unnecessary for the genome-wide genotype data considered here, although they may be quite reasonable for the problems considered in other contexts. Existing methods typically estimate Model 2 by calculating the joint posterior distribution of the $h_{kj}$ based on an assumed prior distribution of the latent variables.  

Our LFA approach for estimating the row basis of $\bf L$ is nonparametric since it does not require a prior assumption on the distribution of latent variables, $\bH$. The model fitting methods of ref. \cn{BKM2011} are too computationally intensive for high-dimensional data, requiring many iterations and potential convergence issues. Our proposed algorithm requires performing SVD twice, which leads to a dramatic reduction in computational burden and difficulties.  Engelhardt and Stephens (2010) \cn{Engelhardt2010} make an interesting connection between classical factor analysis models of $\bF$ and other models of population structure, but the factor analysis model runs into the difficulty that the latent factors are assumed to be Normal distributed, and the constraint that alleles frequencies are in $[0,1]$ is not easily accommodated by this continuous, real-valued model.  
  
Several extensions of PCA to categorical data have been proposed \cn{Schapire2002,Schein2003,Guo2008}. We found that the algorithms perform very slowly on genome-wide genotyping data, and the estimation can be quite poor when $d > 1$.  Also, PCA is essentially a method for characterizing variance in data \cn{Jolliffe10}, and the latent variable approach is more directly aimed at uncovering latent population structure. Non-negative matrix factorization (NMF) \cn{Paatero1994} is another matrix factorization for count data (e.g., Poisson random variables). This identifies two non-negative matrices whose product approximates the original matrix. However, similarly to PCA, we do not find that this approach easily translates into interpretable models of population and it is computationally intensive.  NMF has proven to be quite useful as a numerical tool for decomposing images into parts humans recognize as distinct \cn{Lee1999}. 

\clearpage
\section{{\sc Supplementary Figures and Tables}}  
\begin{figure}[!h]
\centering
\includegraphics[width=0.7\textwidth]{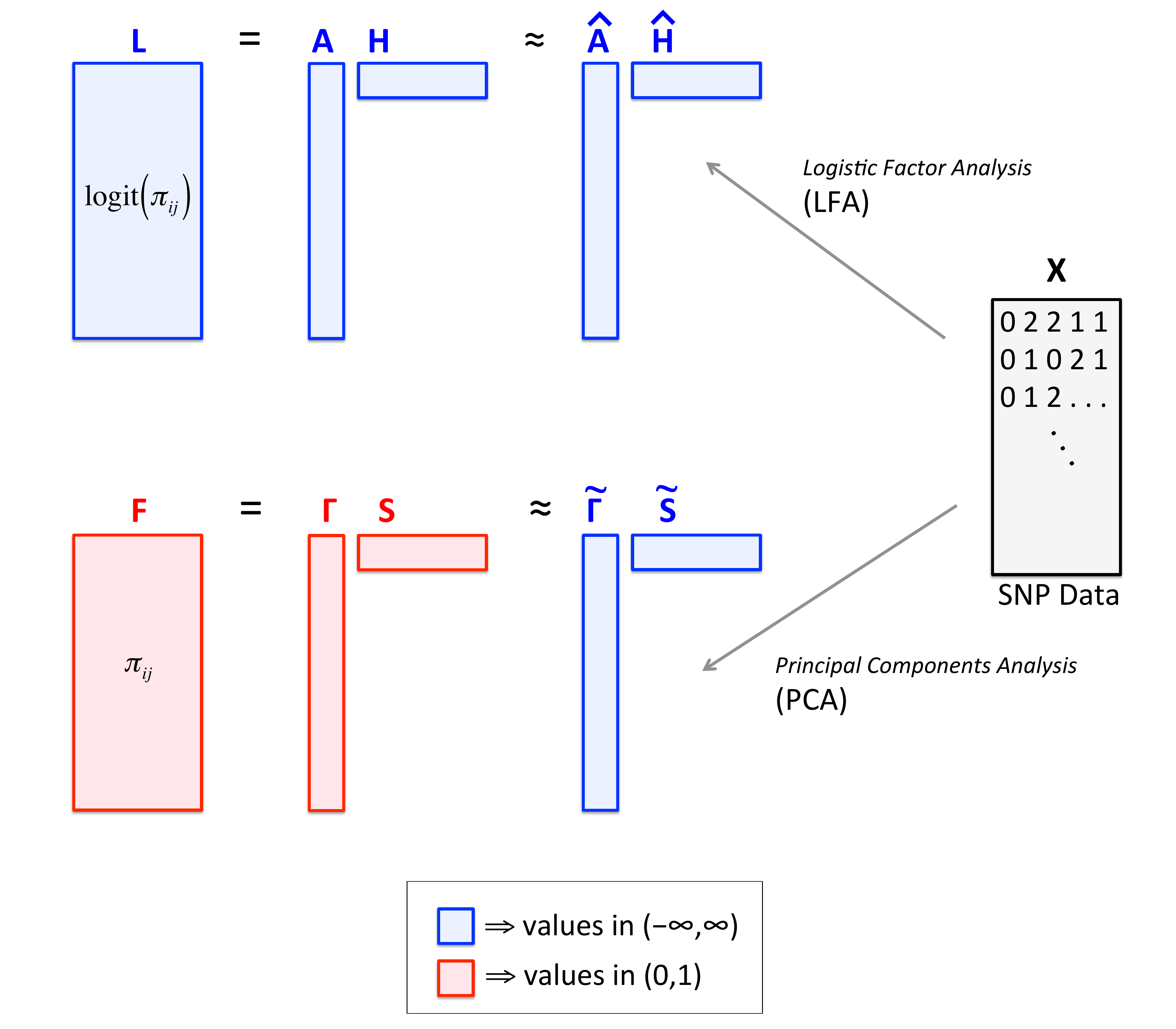}
\caption{A comparison of LFA model (2) and its estimate to model (1) and its PCA estimate.  The proposed LFA approach first models the logit of the individual-specific allele frequencies in terms of the product of two matrices, the left matrix establishing how population structure is present in allele frequencies, and the right matrix giving the structure.  Whereas the LFA approach preserves the scale of the model through the estimate (all real-valued numbers), the same is not true to PCA.  This leads to issues in the estimation of individual-specific allele frequencies when utilizing PCA.  We have shown, however, that PCA estimates very well a row basis for $\bS$ from Model 1.  This connects PCA to an explicit model of population structure.}
\label{fig:decomp}
\end{figure}

\clearpage

\clearpage


\clearpage
\begin{sidewaystable}
\caption{Accuracy in estimating $\pi_{ij}$ parameters by the PCA based method and LFA. Each row is a different simulation scenario. Each column is the accuracy of a method's fits with the given metric.} 
\label{tab:err}
\begin{tabular*}{\linewidth}{ @{\extracolsep{\fill}} ll |cccc|cccc|cccc @{}}
\toprule
\multicolumn{2}{c}{Scenario} & \multicolumn{4}{c}{Median KL} &
\multicolumn{4}{c}{Mean Abs. Err.} & \multicolumn{4}{c}{RMSE}  \\
\midrule \midrule \addlinespace \\
 & & PCA & LFA & ADX & FS & PCA & LFA & ADX & FS & PCA & LFA & ADX & FS \\
\cmidrule{3-6} \cmidrule{7-10} \cmidrule{11-14}
\addlinespace \\
\multirow{1}{*}{\rotatebox{90}{BN}} 
&  & 6.9\e{-5} & 6.8\e{-5} & 2.6\e{-3} & 2.6\e{-3}   & 5.8\e{-3} & 5.8\e{-3} & 3.7\e{-2} & 3.7\e{-2}   & 7.5\e{-3} & 7.5\e{-3} & 5.8\e{-2} & 5.8\e{-2} \\
\midrule \addlinespace \\
\multirow{4}{*}{\rotatebox{90}{PSD}} 
& $\alpha=0.01$  & 7.0\e{-5} & 7.3\e{-5} & 1.6\e{-2} & 1.6\e{-2}   & 5.6\e{-3} & 5.8\e{-3} & 9.7\e{-2} & 9.7\e{-2}   & 7.2\e{-3} & 7.6\e{-3} & 1.7\e{-1} & 1.7\e{-1} \\
& $\alpha=0.1$   & 6.7\e{-5} & 9.2\e{-5} & 3.6\e{-2} & 3.6\e{-2}   & 5.6\e{-3} & 6.9\e{-3} & 1.6\e{-1} & 1.6\e{-1}   & 7.2\e{-3} & 9.3\e{-3} & 2.4\e{-1} & 2.4\e{-1} \\
& $\alpha=0.5$   & 6.3\e{-5} & 8.5\e{-5} & 5.4\e{-2} & 5.4\e{-2}   & 5.6\e{-3} & 6.8\e{-3} & 1.4\e{-1} & 1.4\e{-1}   & 7.3\e{-3} & 9.0\e{-3} & 1.8\e{-1} & 1.8\e{-1} \\
& $\alpha=1.0$   & 6.1\e{-5} & 7.4\e{-5} & 3.3\e{-2} & 3.3\e{-2}   & 5.6\e{-3} & 6.3\e{-3} & 1.4\e{-1} & 1.4\e{-1}   & 7.4\e{-3} & 8.4\e{-3} & 2.2\e{-1} & 2.2\e{-1} \\
\midrule \addlinespace \\
\multirow{4}{*}{\rotatebox{90}{Spatial}} 
& $a=0.1$   & 7.3\e{-5} & 1.2\e{-4} & 8.2\e{-3} & 8.1\e{-3}   & 5.5\e{-3} & 7.6\e{-3} & 7.4\e{-2} & 7.4\e{-2}   & 7.0\e{-3} & 1.0\e{-2} & 1.2\e{-1} & 1.2\e{-1} \\
& $a=0.25$  & 6.9\e{-5} & 1.1\e{-4} & 8.6\e{-3} & 8.6\e{-3}   & 5.6\e{-3} & 7.4\e{-3} & 9.3\e{-2} & 9.3\e{-2}   & 7.2\e{-3} & 9.8\e{-3} & 1.6\e{-1} & 1.6\e{-1} \\
& $a=0.5$   & 6.6\e{-5} & 9.5\e{-5} & 1.0\e{-2} & 1.0\e{-2}   & 5.6\e{-3} & 6.9\e{-3} & 6.7\e{-2} & 6.7\e{-2}   & 7.2\e{-3} & 9.2\e{-3} & 1.0\e{-1} & 1.0\e{-1} \\
& $a=1.0$   & 6.3\e{-5} & 7.8\e{-5} & 1.2\e{-2} & 1.2\e{-2}   & 5.7\e{-3} & 6.4\e{-3} & 1.1\e{-1} & 1.1\e{-1}   & 7.4\e{-3} & 8.5\e{-3} & 1.7\e{-1} & 1.7\e{-1} \\
\midrule \addlinespace \\
\multirow{4}{*}{\rotatebox{90}{TGP fit}} 
& PCA     & 4.1\e{-4} & 5.2\e{-4} & 2.8\e{-3} & 3.4\e{-3}   & 1.3\e{-2} & 1.5\e{-2} & 8.1\e{-2} & 8.3\e{-2}   & 1.8\e{-2} & 2.1\e{-2} & 1.5\e{-1} & 1.5\e{-1} \\
& LFA     & 4.3\e{-4} & 4.8\e{-4} & 2.4\e{-3} & 2.7\e{-3}   & 1.3\e{-2} & 1.4\e{-2} & 7.9\e{-2} & 8.1\e{-2}   & 1.8\e{-2} & 2.0\e{-2} & 1.4\e{-1} & 1.5\e{-1} \\
& ADX     & 5.4\e{-4} & 4.4\e{-4} & 5.0\e{-3} & 5.5\e{-3}   & 1.5\e{-2} & 1.3\e{-2} & 1.1\e{-1} & 1.1\e{-1}   & 2.0\e{-2} & 1.9\e{-2} & 2.0\e{-1} & 2.0\e{-1} \\
& FS      & 4.1\e{-4} & 5.5\e{-4} & 7.8\e{-4} & 9.2\e{-4}   & 1.3\e{-2} & 1.5\e{-2} & 5.6\e{-2} & 5.8\e{-2}   & 1.8\e{-2} & 2.1\e{-2} & 1.3\e{-1} & 1.3\e{-1} \\
\midrule \addlinespace \\
\multirow{4}{*}{\rotatebox{90}{HGDP fit}} 
& PCA     & 1.0\e{-3} & 1.2\e{-3} & 1.3\e{-2} & 1.4\e{-2}   & 2.3\e{-2} & 2.5\e{-2} & 1.2\e{-1} & 1.2\e{-1}   & 3.4\e{-2} & 3.6\e{-2} & 2.2\e{-1} & 2.2\e{-1} \\
& LFA     & 9.9\e{-4} & 1.1\e{-3} & 1.3\e{-2} & 1.2\e{-2}   & 2.2\e{-2} & 2.4\e{-2} & 1.2\e{-1} & 1.2\e{-1}   & 3.5\e{-2} & 3.7\e{-2} & 2.2\e{-1} & 2.2\e{-1} \\
& ADX     & 1.6\e{-3} & 1.4\e{-3} & 2.3\e{-3} & 2.3\e{-3}   & 2.6\e{-2} & 2.6\e{-2} & 5.6\e{-2} & 5.6\e{-2}   & 3.6\e{-2} & 3.7\e{-2} & 1.0\e{-1} & 1.0\e{-1} \\
& FS      & 1.4\e{-3} & 1.6\e{-3} & 3.1\e{-2} & 2.9\e{-2}   & 2.6\e{-2} & 2.7\e{-2} & 1.4\e{-1} & 1.3\e{-1}   & 3.6\e{-2} & 3.8\e{-2} & 2.2\e{-1} & 2.1\e{-1} \\
\bottomrule
\end{tabular*}
\end{sidewaystable}

\clearpage
\pagestyle{empty}
\begin{table}
\vspace{-0.6in}
\caption{\footnotesize The top 50 SNPs most associated with structure in the HGDP data, identified by performing a logistic regression of SNP genotypes on the logistic factors.  Shown are the SNP ID and location, deviance measure of differentiation, gene closest to the SNP, distance to gene (rounded to nearest 10bp), and the variant type (if none shown, then intergenic).}
\label{tab:HGDPtop}
\scriptsize
\begin{center}
\vspace{-0.12in}
\begin{tabular}{rlllrrlrl}
  \hline
 & rsid & chr & position & deviance & genesymbol & locusID & distance & variant type \\ 
  \hline
1 & \href{http://www.ncbi.nlm.nih.gov/projects/SNP/snp_ref.cgi?rs=rs1834640}{rs1834640} & 15 & 48392165 & 1605.28 & SLC24A5 & \href{http://www.ncbi.nlm.nih.gov/gene/283652}{283652} & 21000 &  \\ 
  2 & \href{http://www.ncbi.nlm.nih.gov/projects/SNP/snp_ref.cgi?rs=rs2250072}{rs2250072} & 15 & 48384907 & 1313.82 & SLC24A5 & \href{http://www.ncbi.nlm.nih.gov/gene/283652}{283652} & 28260 &  \\ 
  3 & \href{http://www.ncbi.nlm.nih.gov/projects/SNP/snp_ref.cgi?rs=rs12440301}{rs12440301} & 15 & 48389924 & 1263.83 & SLC24A5 & \href{http://www.ncbi.nlm.nih.gov/gene/283652}{283652} & 23240 &  \\ 
  4 & \href{http://www.ncbi.nlm.nih.gov/projects/SNP/snp_ref.cgi?rs=rs260690}{rs260690} & 2 & 109579738 & 1262.72 & EDAR & \href{http://www.ncbi.nlm.nih.gov/gene/10913}{10913} &   0 & intron-variant \\ 
  5 & \href{http://www.ncbi.nlm.nih.gov/projects/SNP/snp_ref.cgi?rs=rs9837708}{rs9837708} & 3 & 71487582 & 1189.48 & FOXP1 & \href{http://www.ncbi.nlm.nih.gov/gene/27086}{27086} &   0 & intron-variant \\ 
  6 & \href{http://www.ncbi.nlm.nih.gov/projects/SNP/snp_ref.cgi?rs=rs260714}{rs260714} & 2 & 109562495 & 1184.50 & EDAR & \href{http://www.ncbi.nlm.nih.gov/gene/10913}{10913} &   0 & intron-variant \\ 
  7 & \href{http://www.ncbi.nlm.nih.gov/projects/SNP/snp_ref.cgi?rs=rs4918664}{rs4918664} & 10 & 94921065 & 1178.40 & XRCC6P1 & \href{http://www.ncbi.nlm.nih.gov/gene/387703}{387703} & 45340 &  \\ 
  8 & \href{http://www.ncbi.nlm.nih.gov/projects/SNP/snp_ref.cgi?rs=rs10882168}{rs10882168} & 10 & 94929434 & 1160.99 & XRCC6P1 & \href{http://www.ncbi.nlm.nih.gov/gene/387703}{387703} & 36970 &  \\ 
  9 & \href{http://www.ncbi.nlm.nih.gov/projects/SNP/snp_ref.cgi?rs=rs300153}{rs300153} & 2 & 17986417 & 1143.48 & MSGN1 & \href{http://www.ncbi.nlm.nih.gov/gene/343930}{343930} & 11360 &  \\ 
  10 & \href{http://www.ncbi.nlm.nih.gov/projects/SNP/snp_ref.cgi?rs=rs9809818}{rs9809818} & 3 & 71480566 & 1135.58 & FOXP1 & \href{http://www.ncbi.nlm.nih.gov/gene/27086}{27086} &   0 & intron-variant \\ 
  11 & \href{http://www.ncbi.nlm.nih.gov/projects/SNP/snp_ref.cgi?rs=rs6583859}{rs6583859} & 10 & 94893473 & 1119.25 & NIP7P1 & \href{http://www.ncbi.nlm.nih.gov/gene/389997}{389997} & 26290 &  \\ 
  12 & \href{http://www.ncbi.nlm.nih.gov/projects/SNP/snp_ref.cgi?rs=rs11187300}{rs11187300} & 10 & 94920291 & 1114.22 & XRCC6P1 & \href{http://www.ncbi.nlm.nih.gov/gene/387703}{387703} & 46120 &  \\ 
  13 & \href{http://www.ncbi.nlm.nih.gov/projects/SNP/snp_ref.cgi?rs=rs260698}{rs260698} & 2 & 109566759 & 1111.64 & EDAR & \href{http://www.ncbi.nlm.nih.gov/gene/10913}{10913} &   0 & intron-variant \\ 
  14 & \href{http://www.ncbi.nlm.nih.gov/projects/SNP/snp_ref.cgi?rs=rs1834619}{rs1834619} & 2 & 17901485 & 1111.40 & SMC6 & \href{http://www.ncbi.nlm.nih.gov/gene/79677}{79677} &   0 & intron-variant \\ 
  15 & \href{http://www.ncbi.nlm.nih.gov/projects/SNP/snp_ref.cgi?rs=rs11637235}{rs11637235} & 15 & 48633153 & 1104.45 & DUT & \href{http://www.ncbi.nlm.nih.gov/gene/1854}{1854} &   0 & intron-variant \\ 
  16 & \href{http://www.ncbi.nlm.nih.gov/projects/SNP/snp_ref.cgi?rs=rs4497887}{rs4497887} & 2 & 125859777 & 1097.13 & RNA5SP102 & \href{http://www.ncbi.nlm.nih.gov/gene/100873373}{100873373} & 169180 &  \\ 
  17 & \href{http://www.ncbi.nlm.nih.gov/projects/SNP/snp_ref.cgi?rs=rs7091054}{rs7091054} & 10 & 95018444 & 1085.45 & RPL17P34 & \href{http://www.ncbi.nlm.nih.gov/gene/643863}{643863} & 25280 &  \\ 
  18 & \href{http://www.ncbi.nlm.nih.gov/projects/SNP/snp_ref.cgi?rs=rs7090105}{rs7090105} & 10 & 75131545 & 1075.50 & ANXA7 & \href{http://www.ncbi.nlm.nih.gov/gene/310}{310} & 3640 &  \\ 
  19 & \href{http://www.ncbi.nlm.nih.gov/projects/SNP/snp_ref.cgi?rs=rs973787}{rs973787} & 4 & 38263893 & 1074.57 & TBC1D1 & \href{http://www.ncbi.nlm.nih.gov/gene/23216}{23216} & 123090 &  \\ 
  20 & \href{http://www.ncbi.nlm.nih.gov/projects/SNP/snp_ref.cgi?rs=rs4279220}{rs4279220} & 4 & 38254182 & 1070.43 & TBC1D1 & \href{http://www.ncbi.nlm.nih.gov/gene/23216}{23216} & 113380 &  \\ 
  21 & \href{http://www.ncbi.nlm.nih.gov/projects/SNP/snp_ref.cgi?rs=rs7556886}{rs7556886} & 2 & 17908130 & 1062.58 & SMC6 & \href{http://www.ncbi.nlm.nih.gov/gene/79677}{79677} &   0 & intron-variant \\ 
  22 & \href{http://www.ncbi.nlm.nih.gov/projects/SNP/snp_ref.cgi?rs=rs12473565}{rs12473565} & 2 & 175163335 & 1056.31 & LOC644158 & \href{http://www.ncbi.nlm.nih.gov/gene/644158}{644158} & 1390 &  \\ 
  23 & \href{http://www.ncbi.nlm.nih.gov/projects/SNP/snp_ref.cgi?rs=rs6500380}{rs6500380} & 16 & 48375777 & 1051.10 & LONP2 & \href{http://www.ncbi.nlm.nih.gov/gene/83752}{83752} &   0 & intron-variant \\ 
  24 & \href{http://www.ncbi.nlm.nih.gov/projects/SNP/snp_ref.cgi?rs=rs2384319}{rs2384319} & 2 & 26206255 & 1033.88 & KIF3C & \href{http://www.ncbi.nlm.nih.gov/gene/3797}{3797} & 810 & upstream-variant-2KB \\ 
  25 & \href{http://www.ncbi.nlm.nih.gov/projects/SNP/snp_ref.cgi?rs=rs12220128}{rs12220128} & 10 & 94975011 & 1023.79 & XRCC6P1 & \href{http://www.ncbi.nlm.nih.gov/gene/387703}{387703} & 6090 &  \\ 
  26 & \href{http://www.ncbi.nlm.nih.gov/projects/SNP/snp_ref.cgi?rs=rs17034770}{rs17034770} & 2 & 109616376 & 1019.03 & EDAR & \href{http://www.ncbi.nlm.nih.gov/gene/10913}{10913} & 10540 &  \\ 
  27 & \href{http://www.ncbi.nlm.nih.gov/projects/SNP/snp_ref.cgi?rs=rs3792006}{rs3792006} & 2 & 26498222 & 998.96 & HADHB & \href{http://www.ncbi.nlm.nih.gov/gene/3032}{3032} &   0 & intron-variant \\ 
  28 & \href{http://www.ncbi.nlm.nih.gov/projects/SNP/snp_ref.cgi?rs=rs4918924}{rs4918924} & 10 & 94976956 & 994.79 & XRCC6P1 & \href{http://www.ncbi.nlm.nih.gov/gene/387703}{387703} & 8030 &  \\ 
  29 & \href{http://www.ncbi.nlm.nih.gov/projects/SNP/snp_ref.cgi?rs=rs1984996}{rs1984996} & 10 & 95008745 & 990.92 & RPL17P34 & \href{http://www.ncbi.nlm.nih.gov/gene/643863}{643863} & 34980 &  \\ 
  30 & \href{http://www.ncbi.nlm.nih.gov/projects/SNP/snp_ref.cgi?rs=rs3751631}{rs3751631} & 15 & 52534344 & 987.33 & MYO5C & \href{http://www.ncbi.nlm.nih.gov/gene/55930}{55930} &   0 & reference,synonymous-codon \\ 
  31 & \href{http://www.ncbi.nlm.nih.gov/projects/SNP/snp_ref.cgi?rs=rs4578856}{rs4578856} & 2 & 17853388 & 987.29 & SMC6 & \href{http://www.ncbi.nlm.nih.gov/gene/79677}{79677} &   0 & intron-variant \\ 
  32 & \href{http://www.ncbi.nlm.nih.gov/projects/SNP/snp_ref.cgi?rs=rs13397666}{rs13397666} & 2 & 109544052 & 986.80 & EDAR & \href{http://www.ncbi.nlm.nih.gov/gene/10913}{10913} &   0 & intron-variant \\ 
  33 & \href{http://www.ncbi.nlm.nih.gov/projects/SNP/snp_ref.cgi?rs=rs12619554}{rs12619554} & 2 & 17352372 & 986.20 & ZFYVE9P2 & \href{http://www.ncbi.nlm.nih.gov/gene/100420972}{100420972} & 113180 &  \\ 
  34 & \href{http://www.ncbi.nlm.nih.gov/projects/SNP/snp_ref.cgi?rs=rs3736508}{rs3736508} & 11 & 45975130 & 981.05 & PHF21A & \href{http://www.ncbi.nlm.nih.gov/gene/51317}{51317} &   0 & missense,reference \\ 
  35 & \href{http://www.ncbi.nlm.nih.gov/projects/SNP/snp_ref.cgi?rs=rs12472075}{rs12472075} & 2 & 177691130 & 973.02 & RPL29P8 & \href{http://www.ncbi.nlm.nih.gov/gene/100131991}{100131991} & 16650 &  \\ 
  36 & \href{http://www.ncbi.nlm.nih.gov/projects/SNP/snp_ref.cgi?rs=rs9522149}{rs9522149} & 13 & 111827167 & 965.50 & ARHGEF7 & \href{http://www.ncbi.nlm.nih.gov/gene/8874}{8874} &   0 & intron-variant \\ 
  37 & \href{http://www.ncbi.nlm.nih.gov/projects/SNP/snp_ref.cgi?rs=rs2917454}{rs2917454} & 10 & 78892415 & 964.40 & KCNMA1 & \href{http://www.ncbi.nlm.nih.gov/gene/3778}{3778} &   0 & intron-variant \\ 
  38 & \href{http://www.ncbi.nlm.nih.gov/projects/SNP/snp_ref.cgi?rs=rs10882183}{rs10882183} & 10 & 94974083 & 961.04 & XRCC6P1 & \href{http://www.ncbi.nlm.nih.gov/gene/387703}{387703} & 5160 &  \\ 
  39 & \href{http://www.ncbi.nlm.nih.gov/projects/SNP/snp_ref.cgi?rs=rs10079352}{rs10079352} & 5 & 117494640 & 960.33 & LOC100505811 & \href{http://www.ncbi.nlm.nih.gov/gene/100505811}{100505811} & 123620 &  \\ 
  40 & \href{http://www.ncbi.nlm.nih.gov/projects/SNP/snp_ref.cgi?rs=rs10935320}{rs10935320} & 3 & 139056584 & 958.33 & MRPS22 & \href{http://www.ncbi.nlm.nih.gov/gene/56945}{56945} & 6270 &  \\ 
  41 & \href{http://www.ncbi.nlm.nih.gov/projects/SNP/snp_ref.cgi?rs=rs9571407}{rs9571407} & 13 & 34886039 & 957.04 & LINC00457 & \href{http://www.ncbi.nlm.nih.gov/gene/100874179}{100874179} & 123540 &  \\ 
  42 & \href{http://www.ncbi.nlm.nih.gov/projects/SNP/snp_ref.cgi?rs=rs6542787}{rs6542787} & 2 & 109556365 & 955.56 & EDAR & \href{http://www.ncbi.nlm.nih.gov/gene/10913}{10913} &   0 & intron-variant \\ 
  43 & \href{http://www.ncbi.nlm.nih.gov/projects/SNP/snp_ref.cgi?rs=rs953035}{rs953035} & 1 & 36079508 & 954.67 & PSMB2 & \href{http://www.ncbi.nlm.nih.gov/gene/5690}{5690} &   0 & intron-variant \\ 
  44 & \href{http://www.ncbi.nlm.nih.gov/projects/SNP/snp_ref.cgi?rs=rs4657449}{rs4657449} & 1 & 165465281 & 951.72 & LOC400794 & \href{http://www.ncbi.nlm.nih.gov/gene/400794}{400794} &   0 & intron-variant \\ 
  45 & \href{http://www.ncbi.nlm.nih.gov/projects/SNP/snp_ref.cgi?rs=rs9960403}{rs9960403} & 18 & 13437993 & 949.43 & LDLRAD4 & \href{http://www.ncbi.nlm.nih.gov/gene/753}{753} &   0 & intron-variant \\ 
  46 & \href{http://www.ncbi.nlm.nih.gov/projects/SNP/snp_ref.cgi?rs=rs203150}{rs203150} & 18 & 38037221 & 944.32 & RPL17P45 & \href{http://www.ncbi.nlm.nih.gov/gene/100271414}{100271414} & 312750 &  \\ 
  47 & \href{http://www.ncbi.nlm.nih.gov/projects/SNP/snp_ref.cgi?rs=rs2823882}{rs2823882} & 21 & 17934419 & 942.05 & LINC00478 & \href{http://www.ncbi.nlm.nih.gov/gene/388815}{388815} &   0 & intron-variant \\ 
  48 & \href{http://www.ncbi.nlm.nih.gov/projects/SNP/snp_ref.cgi?rs=rs10886189}{rs10886189} & 10 & 119753963 & 937.81 & RAB11FIP2 & \href{http://www.ncbi.nlm.nih.gov/gene/22841}{22841} & 10460 &  \\ 
  49 & \href{http://www.ncbi.nlm.nih.gov/projects/SNP/snp_ref.cgi?rs=rs2441727}{rs2441727} & 10 & 68224886 & 937.08 & CTNNA3 & \href{http://www.ncbi.nlm.nih.gov/gene/29119}{29119} &   0 & intron-variant \\ 
  50 & \href{http://www.ncbi.nlm.nih.gov/projects/SNP/snp_ref.cgi?rs=rs310644}{rs310644} & 20 & 62159504 & 931.90 & PTK6 & \href{http://www.ncbi.nlm.nih.gov/gene/5753}{5753} & 260 & downstream-variant-500B \\ 
   \hline
\hline
\end{tabular}
\end{center}
\normalsize
\end{table}

\clearpage
\begin{table}
\vspace{-0.6in}
\caption{\footnotesize The top 50 SNPs most associated with structure in the TGP data, identified by performing a logistic regression of SNP genotypes on the logistic factors.  Shown are the SNP ID and location, deviance measure of differentiation, gene closest to the SNP, distance to gene (rounded to nearest 10bp), and the variant type (if none shown, then intergenic).}
\label{tab:TGPtop}
\scriptsize
\begin{center}
\vspace{-0.25in}
\begin{tabular}{rlllrrlrl}
  \hline
 & rsid & chr & position & deviance & genesymbol & locusID & distance & variant type \\ 
  \hline
1 & \href{http://www.ncbi.nlm.nih.gov/projects/SNP/snp_ref.cgi?rs=rs1426654}{rs1426654} & 15 & 48426484 & 3129.76 & SLC24A5 & \href{http://www.ncbi.nlm.nih.gov/gene/283652}{283652} &   0 & missense,reference \\ 
  2 & \href{http://www.ncbi.nlm.nih.gov/projects/SNP/snp_ref.cgi?rs=rs3827760}{rs3827760} & 2 & 109513601 & 2395.27 & EDAR & \href{http://www.ncbi.nlm.nih.gov/gene/10913}{10913} &   0 & missense,reference \\ 
  3 & \href{http://www.ncbi.nlm.nih.gov/projects/SNP/snp_ref.cgi?rs=rs922452}{rs922452} & 2 & 109543883 & 2338.38 & EDAR & \href{http://www.ncbi.nlm.nih.gov/gene/10913}{10913} &   0 & intron-variant \\ 
  4 & \href{http://www.ncbi.nlm.nih.gov/projects/SNP/snp_ref.cgi?rs=rs372985703}{rs372985703} & 17 & 19172196 & 1975.16 & EPN2 & \href{http://www.ncbi.nlm.nih.gov/gene/22905}{22905} &   0 & intron-variant \\ 
  5 & \href{http://www.ncbi.nlm.nih.gov/projects/SNP/snp_ref.cgi?rs=rs4924987}{rs4924987} & 17 & 19247075 & 1949.03 & B9D1 & \href{http://www.ncbi.nlm.nih.gov/gene/27077}{27077} &   0 & intron-variant,missense,reference \\ 
  6 & \href{http://www.ncbi.nlm.nih.gov/projects/SNP/snp_ref.cgi?rs=rs260687}{rs260687} & 2 & 109578855 & 1925.18 & EDAR & \href{http://www.ncbi.nlm.nih.gov/gene/10913}{10913} &   0 & intron-variant \\ 
  7 & \href{http://www.ncbi.nlm.nih.gov/projects/SNP/snp_ref.cgi?rs=rs7209202}{rs7209202} & 17 & 58532239 & 1890.67 & APPBP2 & \href{http://www.ncbi.nlm.nih.gov/gene/10513}{10513} &   0 &  \\ 
  8 & \href{http://www.ncbi.nlm.nih.gov/projects/SNP/snp_ref.cgi?rs=rs7211872}{rs7211872} & 17 & 58550725 & 1890.67 & APPBP2 & \href{http://www.ncbi.nlm.nih.gov/gene/10513}{10513} &   0 &  \\ 
  9 & \href{http://www.ncbi.nlm.nih.gov/projects/SNP/snp_ref.cgi?rs=rs67929453}{rs67929453} & 3 & 139109825 & 1890.57 & LOC100507291 & \href{http://www.ncbi.nlm.nih.gov/gene/100507291}{100507291} &   0 & intron-variant,upstream-variant-2KB \\ 
  10 & \href{http://www.ncbi.nlm.nih.gov/projects/SNP/snp_ref.cgi?rs=rs260643}{rs260643} & 2 & 109539653 & 1850.71 & EDAR & \href{http://www.ncbi.nlm.nih.gov/gene/10913}{10913} &   0 & intron-variant \\ 
  11 & \href{http://www.ncbi.nlm.nih.gov/projects/SNP/snp_ref.cgi?rs=rs260707}{rs260707} & 2 & 109574150 & 1838.37 & EDAR & \href{http://www.ncbi.nlm.nih.gov/gene/10913}{10913} &   0 & intron-variant \\ 
  12 & \href{http://www.ncbi.nlm.nih.gov/projects/SNP/snp_ref.cgi?rs=rs1545071}{rs1545071} & 18 & 67695505 & 1821.35 & RTTN & \href{http://www.ncbi.nlm.nih.gov/gene/25914}{25914} &   0 & intron-variant \\ 
  13 & \href{http://www.ncbi.nlm.nih.gov/projects/SNP/snp_ref.cgi?rs=rs12729599}{rs12729599} & 1 & 1323078 & 1812.91 & CCNL2 & \href{http://www.ncbi.nlm.nih.gov/gene/81669}{81669} &   0 & intron-variant \\ 
  14 & \href{http://www.ncbi.nlm.nih.gov/projects/SNP/snp_ref.cgi?rs=rs12347078}{rs12347078} & 9 & 344508 & 1811.16 & DOCK8 & \href{http://www.ncbi.nlm.nih.gov/gene/81704}{81704} &   0 & intron-variant \\ 
  15 & \href{http://www.ncbi.nlm.nih.gov/projects/SNP/snp_ref.cgi?rs=rs12142199}{rs12142199} & 1 & 1249187 & 1779.28 & CPSF3L & \href{http://www.ncbi.nlm.nih.gov/gene/54973}{54973} &   0 & reference,synonymous-codon \\ 
  16 & \href{http://www.ncbi.nlm.nih.gov/projects/SNP/snp_ref.cgi?rs=rs12953952}{rs12953952} & 18 & 67737927 & 1750.15 & RTTN & \href{http://www.ncbi.nlm.nih.gov/gene/25914}{25914} &   0 & intron-variant \\ 
  17 & \href{http://www.ncbi.nlm.nih.gov/projects/SNP/snp_ref.cgi?rs=rs9467091}{rs9467091} & 6 & 10651772 & 1746.75 & GCNT6 & \href{http://www.ncbi.nlm.nih.gov/gene/644378}{644378} & 4270 &  \\ 
  18 & \href{http://www.ncbi.nlm.nih.gov/projects/SNP/snp_ref.cgi?rs=rs7165971}{rs7165971} & 15 & 55921013 & 1736.83 & PRTG & \href{http://www.ncbi.nlm.nih.gov/gene/283659}{283659} &   0 & intron-variant \\ 
  19 & \href{http://www.ncbi.nlm.nih.gov/projects/SNP/snp_ref.cgi?rs=rs6132532}{rs6132532} & 20 & 2315543 & 1730.64 & TGM3 & \href{http://www.ncbi.nlm.nih.gov/gene/7053}{7053} &   0 & intron-variant \\ 
  20 & \href{http://www.ncbi.nlm.nih.gov/projects/SNP/snp_ref.cgi?rs=rs959071}{rs959071} & 17 & 19142226 & 1729.18 & EPN2 & \href{http://www.ncbi.nlm.nih.gov/gene/22905}{22905} &   0 & intron-variant \\ 
  21 & \href{http://www.ncbi.nlm.nih.gov/projects/SNP/snp_ref.cgi?rs=rs10962599}{rs10962599} & 9 & 16795286 & 1726.24 & BNC2 & \href{http://www.ncbi.nlm.nih.gov/gene/54796}{54796} &   0 & intron-variant \\ 
  22 & \href{http://www.ncbi.nlm.nih.gov/projects/SNP/snp_ref.cgi?rs=rs967377}{rs967377} & 20 & 53222217 & 1724.93 & DOK5 & \href{http://www.ncbi.nlm.nih.gov/gene/55816}{55816} &   0 & intron-variant \\ 
  23 & \href{http://www.ncbi.nlm.nih.gov/projects/SNP/snp_ref.cgi?rs=rs4891381}{rs4891381} & 18 & 67595449 & 1723.79 & CD226 & \href{http://www.ncbi.nlm.nih.gov/gene/10666}{10666} &   0 & intron-variant \\ 
  24 & \href{http://www.ncbi.nlm.nih.gov/projects/SNP/snp_ref.cgi?rs=rs377561427}{rs377561427} & 15 & 63988357 & 1713.98 & HERC1 & \href{http://www.ncbi.nlm.nih.gov/gene/8925}{8925} &   0 & frameshift-variant,reference \\ 
  25 & \href{http://www.ncbi.nlm.nih.gov/projects/SNP/snp_ref.cgi?rs=rs73889254}{rs73889254} & 22 & 46762214 & 1711.40 & CELSR1 & \href{http://www.ncbi.nlm.nih.gov/gene/9620}{9620} &   0 & intron-variant \\ 
  26 & \href{http://www.ncbi.nlm.nih.gov/projects/SNP/snp_ref.cgi?rs=rs4918664}{rs4918664} & 10 & 94921065 & 1700.64 & XRCC6P1 & \href{http://www.ncbi.nlm.nih.gov/gene/387703}{387703} & 45340 &  \\ 
  27 & \href{http://www.ncbi.nlm.nih.gov/projects/SNP/snp_ref.cgi?rs=rs2759281}{rs2759281} & 1 & 204866365 & 1691.03 & NFASC & \href{http://www.ncbi.nlm.nih.gov/gene/23114}{23114} &   0 & intron-variant \\ 
  28 & \href{http://www.ncbi.nlm.nih.gov/projects/SNP/snp_ref.cgi?rs=rs12065033}{rs12065033} & 1 & 173579034 & 1682.54 & ANKRD45 & \href{http://www.ncbi.nlm.nih.gov/gene/339416}{339416} &   0 & utr-variant-3-prime \\ 
  29 & \href{http://www.ncbi.nlm.nih.gov/projects/SNP/snp_ref.cgi?rs=rs9796793}{rs9796793} & 16 & 30495652 & 1681.28 & ITGAL & \href{http://www.ncbi.nlm.nih.gov/gene/3683}{3683} &   0 & intron-variant \\ 
  30 & \href{http://www.ncbi.nlm.nih.gov/projects/SNP/snp_ref.cgi?rs=rs1240708}{rs1240708} & 1 & 1335790 & 1675.48 & LOC148413 & \href{http://www.ncbi.nlm.nih.gov/gene/148413}{148413} &   0 & intron-variant,upstream-variant-2KB \\ 
  31 & \href{http://www.ncbi.nlm.nih.gov/projects/SNP/snp_ref.cgi?rs=rs2615876}{rs2615876} & 10 & 117665860 & 1670.53 & ATRNL1 & \href{http://www.ncbi.nlm.nih.gov/gene/26033}{26033} &   0 & intron-variant \\ 
  32 & \href{http://www.ncbi.nlm.nih.gov/projects/SNP/snp_ref.cgi?rs=rs2823882}{rs2823882} & 21 & 17934419 & 1669.32 & LINC00478 & \href{http://www.ncbi.nlm.nih.gov/gene/388815}{388815} &   0 & intron-variant \\ 
  33 & \href{http://www.ncbi.nlm.nih.gov/projects/SNP/snp_ref.cgi?rs=rs8097206}{rs8097206} & 18 & 38024931 & 1663.29 & RPL17P45 & \href{http://www.ncbi.nlm.nih.gov/gene/100271414}{100271414} & 300460 &  \\ 
  34 & \href{http://www.ncbi.nlm.nih.gov/projects/SNP/snp_ref.cgi?rs=rs8071181}{rs8071181} & 17 & 58508582 & 1662.44 & C17orf64 & \href{http://www.ncbi.nlm.nih.gov/gene/124773}{124773} &   0 & reference,synonymous-codon \\ 
  35 & \href{http://www.ncbi.nlm.nih.gov/projects/SNP/snp_ref.cgi?rs=rs1075389}{rs1075389} & 15 & 64174177 & 1661.21 & MIR422A & \href{http://www.ncbi.nlm.nih.gov/gene/494334}{494334} & 10950 &  \\ 
  36 & \href{http://www.ncbi.nlm.nih.gov/projects/SNP/snp_ref.cgi?rs=rs6875659}{rs6875659} & 5 & 175158653 & 1657.54 & HRH2 & \href{http://www.ncbi.nlm.nih.gov/gene/3274}{3274} & 22410 &  \\ 
  37 & \href{http://www.ncbi.nlm.nih.gov/projects/SNP/snp_ref.cgi?rs=rs7171940}{rs7171940} & 15 & 64170986 & 1654.01 & MIR422A & \href{http://www.ncbi.nlm.nih.gov/gene/494334}{494334} & 7760 &  \\ 
  38 & \href{http://www.ncbi.nlm.nih.gov/projects/SNP/snp_ref.cgi?rs=rs2148359}{rs2148359} & 9 & 7385508 & 1652.16 & RPL4P5 & \href{http://www.ncbi.nlm.nih.gov/gene/158345}{158345} & 91440 &  \\ 
  39 & \href{http://www.ncbi.nlm.nih.gov/projects/SNP/snp_ref.cgi?rs=rs7531501}{rs7531501} & 1 & 234338303 & 1648.15 & SLC35F3 & \href{http://www.ncbi.nlm.nih.gov/gene/148641}{148641} &   0 & intron-variant \\ 
  40 & \href{http://www.ncbi.nlm.nih.gov/projects/SNP/snp_ref.cgi?rs=rs57742857}{rs57742857} & 15 & 93567352 & 1645.21 & CHD2 & \href{http://www.ncbi.nlm.nih.gov/gene/1106}{1106} &   0 & intron-variant \\ 
  41 & \href{http://www.ncbi.nlm.nih.gov/projects/SNP/snp_ref.cgi?rs=rs931564}{rs931564} & 17 & 58631702 & 1636.86 & LOC388406 & \href{http://www.ncbi.nlm.nih.gov/gene/388406}{388406} & 10200 &  \\ 
  42 & \href{http://www.ncbi.nlm.nih.gov/projects/SNP/snp_ref.cgi?rs=rs4738296}{rs4738296} & 8 & 73857539 & 1632.70 & LOC100288310 & \href{http://www.ncbi.nlm.nih.gov/gene/100288310}{100288310} &   0 & intron-variant \\ 
  43 & \href{http://www.ncbi.nlm.nih.gov/projects/SNP/snp_ref.cgi?rs=rs4402785}{rs4402785} & 2 & 104766351 & 1631.33 & LOC100287010 & \href{http://www.ncbi.nlm.nih.gov/gene/100287010}{100287010} & 228950 &  \\ 
  44 & \href{http://www.ncbi.nlm.nih.gov/projects/SNP/snp_ref.cgi?rs=rs12988506}{rs12988506} & 2 & 33162854 & 1630.14 & LOC100271832 & \href{http://www.ncbi.nlm.nih.gov/gene/100271832}{100271832} &   0 & intron-variant \\ 
  45 & \href{http://www.ncbi.nlm.nih.gov/projects/SNP/snp_ref.cgi?rs=rs9410664}{rs9410664} & 9 & 91196828 & 1625.48 & NXNL2 & \href{http://www.ncbi.nlm.nih.gov/gene/158046}{158046} & 6120 &  \\ 
  46 & \href{http://www.ncbi.nlm.nih.gov/projects/SNP/snp_ref.cgi?rs=rs2041564}{rs2041564} & 2 & 72453847 & 1623.91 & EXOC6B & \href{http://www.ncbi.nlm.nih.gov/gene/23233}{23233} &   0 & intron-variant \\ 
  47 & \href{http://www.ncbi.nlm.nih.gov/projects/SNP/snp_ref.cgi?rs=rs6024103}{rs6024103} & 20 & 54034601 & 1623.41 & LOC101927796 & \href{http://www.ncbi.nlm.nih.gov/gene/101927796}{101927796} & 2270 &  \\ 
  48 & \href{http://www.ncbi.nlm.nih.gov/projects/SNP/snp_ref.cgi?rs=rs6583859}{rs6583859} & 10 & 94893473 & 1619.79 & NIP7P1 & \href{http://www.ncbi.nlm.nih.gov/gene/389997}{389997} & 26290 &  \\ 
  49 & \href{http://www.ncbi.nlm.nih.gov/projects/SNP/snp_ref.cgi?rs=rs12913832}{rs12913832} & 15 & 28365618 & 1611.23 & HERC2 & \href{http://www.ncbi.nlm.nih.gov/gene/8924}{8924} &   0 & intron-variant \\ 
  50 & \href{http://www.ncbi.nlm.nih.gov/projects/SNP/snp_ref.cgi?rs=rs632876}{rs632876} & 2 & 216572452 & 1610.26 & LINC00607 & \href{http://www.ncbi.nlm.nih.gov/gene/646324}{646324} &   0 & intron-variant \\ 
   \hline
\hline
\end{tabular}
\end{center}
\normalsize
 \end{table}

\clearpage
\addcontentsline{toc}{section}{ {\sc References}}
\bibliographystyle{nature}
\bibliography{myrefs}

\end{document}